\documentclass[twocolumn]{aastex63}
\pdfoutput=1 
\usepackage{amsmath,amstext}
\usepackage[T1]{fontenc}
\usepackage{apjfonts} 
\usepackage[figure,figure*]{hypcap}
\usepackage{hyperref}

\shorttitle{Embedded coherent structures from MHD to sub-ion scales at 0.17~AU}
\shortauthors{A. Vinogradov et al.}

\graphicspath{{./}{figures/}}

\begin{document}
\definecolor{darkgreen}{rgb}{0.0, 0.5, 0.0}
\definecolor{orange}{rgb}{1.0, 0.4,0.0}

\newcommand{\sd}[1]{{\color{cyan}{#1}}}
\newcommand{\sdc}[1]{{\color{blue}{Sc: #1}}}
\newcommand{\p}[1]{{\color{magenta}{#1}}}
\newcommand{\pc}[1]{{\color{green}{(P comment: #1)}}}
\newcommand{\pr}[1]{{\color{yellow}{(P remove: #1)}}}
\newcommand{\vl}[1]{{\color{darkgreen}{#1}}}
\newcommand{\vlc}[1]{{\color{orange}{(Vc: #1)}}}
\newcommand{\oa}[1]{#1}
\newcommand{\sv}[1]{{\color{violet}{#1}}} 
\newcommand{\svp}[1]{{\color{blue}{#1}}} 
\newcommand{\svtodo}[1]{{\color{red}{#1}}} 
\newcommand{\correction}[1]{{#1}}

\newcommand{\BE}{\begin{equation}}
\newcommand{\EE}{\end{equation}}
\newcommand{\BA}{\begin{eqnarray}}
\newcommand{\EA}{\end{eqnarray}}
 \renewcommand{\fig}[1]{Fig.~\ref{fig_#1}}
 \newcommand{\figsss}[1]{Figure~\ref{fig_#1}}
 \newcommand{\figs}[2]{Figs.~\ref{fig_#1} and \ref{fig_#2}}
 \newcommand{\figss}[2]{Figs.~\ref{fig_#1} - \ref{fig_#2}}
  \newcommand{\sectlong}[1]{Section~\ref{sect_#1}}
 \newcommand{\sect}[1]{Sect.~\ref{sect_#1}}
 \renewcommand{\ap}[1]{Appendix~\ref{ap_#1}}
 \newcommand{\sects}[2]{Sects.~\ref{sect_#1} and~\ref{sect_#2}}
 \newcommand{\eq}[1]{Eq.~(\ref{eq_#1})}
  \newcommand{\eqp}[1]{(Eq.~\ref{eq_#1})}
 \newcommand{\eqs}[2]{Eqs.~(\ref{eq_#1}) and (\ref{eq_#2})}
 \newcommand{\eqss}[2]{Eqs.~(\ref{eq_#1}) - (\ref{eq_#2})}
 \newcommand{\eqsss}[3]{Eqs.~(\ref{eq_#1}), (\ref{eq_#2}) and (\ref{eq_#3})}
\newcommand{\eg}{{\it e.g.}}
\newcommand{\etal}{{\it et al.}}
\newcommand{\ie}{{\it i.e.}}
\newcommand{\insitu}{{\it in situ }}
\newcommand{\cf}{{\it cf.}}
\newcommand{\tab}[1]{Table~\ref{tab_#1}}

\newcommand{\f}[2]{{\ensuremath{\mathchoice%
        {\dfrac{#1}{#2}}
        {\dfrac{#1}{#2}}
        {\frac{#1}{#2}}
        {\frac{#1}{#2}}
        }}}
\newcommand{\Int}[2]{\ensuremath{\mathchoice%
        {\displaystyle\int_{#1}^{#2}}
        {\displaystyle\int_{#1}^{#2}}
        {\int_{#1}^{#2}}
        {\int_{#1}^{#2}}
        }}

\newcommand{\curl}{ {\bf \nabla} \times}
\renewcommand{\div}[1]{ {\bf \nabla }. #1 }
\newcommand{\grad}{ {\bf \nabla } }
\newcommand{\pder}[2]{\f{\partial #1}{\partial #2}}
\newcommand{\lapO}{ {\nabla^2_\bot } }
\newcommand{\pxy}[2]{\{#1,#2 \}_{x,y}}
\newcommand{\pxeta}[2]{\{#1,#2 \}_{x,\eta}}
\newcommand{\rmd}{{\rm d }}
\newcommand{\tder}[2]{\f{D #1}{D #2}}

\newcommand{\uvec}[1]{\hat{\bf #1}}
\renewcommand{\tild}[1]{\tilde{#1}}
\newcommand{\tA}{\tilde{A}}
\newcommand{\tPhi}{\tilde{\Phi}}

\newcommand{\vA}{\mathbf{A}}
\newcommand{\vB}{\mathbf{B}}
\newcommand{\vj}{\mathbf{j}}
\newcommand{\vv}{\mathbf{v}}

\newcommand{\au}{\rm au}
\newcommand{\degree}{\ensuremath{^\circ}}
\newcommand{\kms}{\rm km~s$^{-1}$}


\newcommand{\eps}{    {Earth, Planets, and Space}}
\newcommand{\adv}{    {Adv. Spa. Res.}}
\newcommand{\annG}{   {Annales Geophysicae}}
\newcommand{\ag}{   {Annales Geophysicae}}
\newcommand{\jastp}{  {J. Atmos. Sol. Terr. Phys.}}
\newcommand{\jfm}{J. Fluid. Mech. }
\newcommand{\lrsp}{    {\it Living Rev. Sol. Phys.}}
\newcommand{\natc}{    {\it Nature Com.}}


\newcommand{\vdag}{(v)^\dagger}
\newcommand\aastex{AAS\TeX}
\newcommand\latex{La\TeX}

\title{Embedded coherent structures from MHD to sub-ion scales in turbulent solar wind at 0.17~AU}

\correspondingauthor{Alexander Vinogradov}
\email{alexander.vinogradov@obspm.fr}

\author[0000-0002-0786-7307]{Alexander Vinogradov}
\affiliation{LESIA, Observatoire de Paris, Université PSL, CNRS, Sorbonne Université, Université de Paris, 5 place Jules Janssen, 92195 Meudon, France}
\affiliation{Space Research Institute of the Russian Academy of Sciences, Moscow, Russia}

\author[0000-0003-3811-2991]{Olga Alexandrova}
\affiliation{LESIA, Observatoire de Paris, Université PSL, CNRS, Sorbonne Université, Université de Paris, 5 place Jules Janssen, 92195 Meudon, France}

\author[0000-0001-8215-6532]{Pascal D\'emoulin}
\affiliation{LESIA, Observatoire de Paris, Université PSL, CNRS, Sorbonne Université, Université de Paris, 5 place Jules Janssen, 92195 Meudon, France}

\author{Anton Artemyev}
\affiliation{Space Research Institute of the Russian Academy of Sciences, Moscow, Russia}
\affiliation{Institute of Geophysics and Planetary Physics, University of California, Los Angeles, CA, USA}

\author{Milan Maksimovic}
\affiliation{LESIA, Observatoire de Paris, Université PSL, CNRS, Sorbonne Université, Université de Paris, 5 place Jules Janssen, 92195 Meudon, France}

\author{André Mangeney}
\affiliation{LESIA, Observatoire de Paris, Université PSL, CNRS, Sorbonne Université, Université de Paris, 5 place Jules Janssen, 92195 Meudon, France}

\author[0000-0002-2008-7647]{Alexei Vasiliev}
\affiliation{Space Research Institute of the Russian Academy of Sciences, Moscow, Russia}

\author{Anatoly Petrukovich}
\affiliation{Space Research Institute of the Russian Academy of Sciences, Moscow, Russia}

\author{Stuart Bale}
\affiliation{Space Science Laboratory, University of California, Berkeley, USA}
\affiliation{Physics department, University of California, Berkeley, CA, USA}

\begin{abstract}
We study intermittent coherent structures in solar wind turbulence from MHD to kinetic plasma scales using Parker Solar Probe data during its first perihelion (at 0.17 au) in the \correction{highly-}Alfv\'enic slow \correction{solar} wind. 
We detect coherent structures 
using Morlet wavelets. 
For the first time, we apply a multi-scale \correction{analysis} in physical space. 
At MHD scales within the inertial range, times scales $\tau\in (1,10^2)$~s, we find (i) current sheets including switchback boundaries and (ii) Alfvén vortices. 
Within these  events are  embedded structures at smaller scales: typically  Alfv\'en vortices at ion scales, \oa{$\tau\in (0.08,1)$~s,} and compressible vortices at sub-ion scales,  \oa{$\tau\in 8( 10^{-3}, 10^{-2})$~s}. The number of coherent structures grows toward smaller scales: we observe 
$\sim200$ events during \correction{a} 5~h time interval at MHD scales, $\sim 10^3$ 
\correction{at} ion scales\correction{,} and $\sim 10^4$ 
at sub-ion scales. 
In general, there are multiple structures of ion and sub-ion scales embedded within one MHD structure. 
There are also examples of ion and sub-ion \correction{scale} structures outside MHD structures.
 
To quantify the relative importance of different \correction{types} of structures, we do a statistical comparison of the observed structures with the expectations of models of the current sheets and vortices. 
The results show the dominance of Alfv\'en vortices at all scales in contrast to the widespread view of \correction{the} dominance of current sheets. This means that Alfv\'en  vortices are 
important  building \correction{blocks} of \correction{Alfv\'enic} solar wind turbulence.

\end{abstract}

\keywords{Solar wind -- space plasma turbulence -- intermittency -- coherent structures}

\section{Introduction} 
Solar wind fluctuations cover a broad range of scales: from macroscopic scales, where the energy is injected into the MHD turbulent cascade, to micro-scales, where kinetic effects play \correction{an} important role, and the energy is dissipated. The dissipation mechanism has not been understood yet.
Numerical simulations indicate that dissipation occurs inhomogeneously \citep{2012Wan,2013Karimabadi,2013Zhdankin,2019Kuzzay}. Regions of increased heating in the solar wind correlate with observations of coherent structures \citep{2011Osman,2013Wu,2015Chasapis,2022Sioulas}. 
Coherent structures can be defined as high-amplitude, stable, localized in space events with phase coherence over its spatial extent \citep{Hussain1986,1988Fiedler,Veltri1999,Bruno2001,Mangeney2001,Farge2015,Alexandrova2020_hdr}.

Different types of coherent structures are observed in the solar wind at different scales. Large-scale flux tubes and flux ropes cover energy-containing scales and the inertial range \citep[e.g.,][]{2000Moldwin,2008Feng,Borovsky2008,2014Janvier,2020Zhao}. 
Current sheets  are usually observed at small scales of the inertial range and at ion scales  \citep[e.g.,][]{1968Siscoe,1969Burlaga,2000salem_phd,2004Knetter,2011Tsurutani,2016Lion,2016Perrone,2019Artemyev}. Recent Solar Orbiter observations reveal embedded ion scale flux rope in a bifurcated current sheet \citep{2021Eastwood}.
Alfv\'en vortices have been identified at \correction{MHD and ion scales} \citep{2003Verkhoglyadova,2016Roberts,2016Lion,2016Perrone,2017Perrone}. \correction{Recent numerical simulation shows that magnetic vortices emerge at the late stage of the reflection-driven turbulence \citep{Meyrand2023}.}
Compressible structures, such as magnetic holes \citep[e.g.,][]{1977Turner,2007Stevens,2020Volwerk}), solitons and shocks \citep{2016Perrone,2000salem_phd} are observed at the end of the inertial range and at ion scales. 

Coherent structures contribute significantly to the magnetic  turbulent spectrum in the solar wind. \cite{Li2011} show that in the presence of current sheets, the inertial range spectrum is closer to the Kolmogorov scaling, $-5/3$, while without current sheets, the spectrum is closer to the Iroshnikov-Kraichnan scaling,  $-3/2$. 
In a case study of a fast wind stream by \cite{2016Lion}, 
the contribution of coherent structures to \correction{the} magnetic field spectrum is up to 40 \% from \correction{the} inertial range down to ion scales.  
Therefore, coherent structures are energetically important elements of solar wind turbulence.

\correction{There are less observations  of coherent structures on sub-ion scales.}
Cluster/STAFF allows to measure sub-ion scale fluctuations at 1~au.
\cite{2016Greco} studied ion \correction{scale} current sheets and showed \correction{the} presence of \correction{many} smaller ones at sub-ion scales. 
\correction{In this study the authors used} the partial variance of increments (PVI) method \citep{greco18}, which is appropriate \correction{for detecting} planar structures \citep[e.g., see the discussion in ][]{2016Lion}. 
Another study with Cluster, but applying Morlet wavelets, shows embedded Alfv\'en vortex\correction{-like} fluctuations at sub-ion scales in a current sheet at ion scales \citep{2018Jovanovic}. \correction{To explain this observation, the} authors \correction{developed} the analytic model of a chain of Alfv\'en vortices embedded in the current sheet. 
\correction{At electron scales there are also signatures of Alfv\'en-vortex like fluctuations detected by two Cluster satellites separated by 7~km only \citep{Alexandrova2020_hdr,Alexandrova2020_arxiv}.}

The first multi-satellite observations of these cylindrical structures \correction{in space plasma were obtained with the Cluster mission in the Earth's magnetosheath behind a quasi-perpendicular bow-shock }\citep{2006Alexandrova,Alexandrova2008NPG}.
Cassini measurements \correction{also} indicate the presence of such structures in the Kronian magnetosheath \citep{2008Alexandrova}.  
Signatures of Alfv\'en vortices in the solar wind  using one satellite have been shown by \citet{2003Verkhoglyadova} and \citet{2016Lion}. 
\cite{2016Roberts} and \cite{2016Perrone,2017Perrone} confirmed the existence of Alfven vortices in the solar wind with 4 satellites of Cluster. \citet{2019Wang} investigated the kinetic effects within an Alfv\'en vortex thanks to MMS measurements in the Earth's magnetosheath.

In the present paper, we study magnetic turbulent fluctuations from MHD inertial range to sub-ion scales with Parker Solar Probe (PSP) data at 0.17~au. 
Using the Morlet wavelet transform, which is a good compromise between time and frequency resolution, we detect intermittent events \correction{that} cover a wide range of scales.
We show that these events correspond to embedded multi-scale structures, from MHD to sub-ion scales. 
Then, we study in more details \correction{the} nature of these structures, which cover the whole turbulent cascade.

The article is organized as follows: 
Section~\ref{sec:data} describes the PSP data used in the analysis.  
\correction{In Section~\ref{sec:spectrum} we identify MHD, ion and sub-ion ranges of scales based on the magnetic spectral properties and positions of ion characteristic scales in the spectrum. }
Section~\ref{sec:detection} is dedicated to the detection method of coherent structures. 
Section~\ref{sec:models} presents several theoretical models of the structures we think we cross by PSP at different scales.
Then\correction{,} we study the sensitivity of minimum variance results for different spacecraft trajectories across the model structures for different noise levels.
In section \ref{sec:structures} we describe \correction{a} few examples of structures \correction{detected simultaneously} at MHD, ion\correction{,} and sub-ion scales.   
Section~\ref{sec:minvar} describes \correction{the} statistical study of the observed coherent events during 5h of the 1st PSP encounter at MHD, ion\correction{,} and sub-ion scales. In section~\ref{sec:conclusion} we summarize and discuss the results.

\section{Data} \label{sec:data}
\begin{figure}[t!]
	\centering
	\includegraphics[width=.99\linewidth]{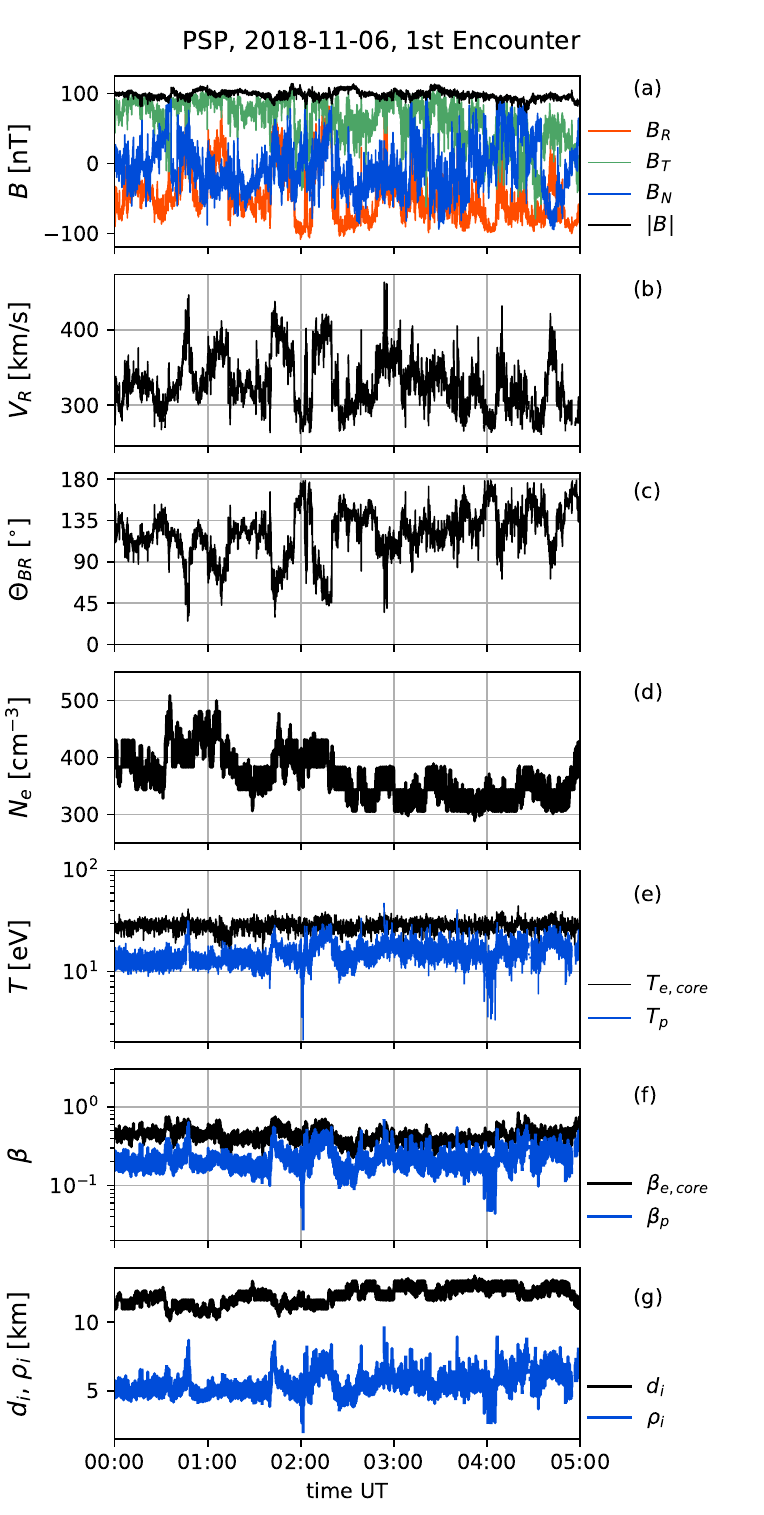}

 \caption{Overview of the solar wind data during the first perihelion of PSP at 0.17~au on November 6, 2018, between 00:00 and 05:00 UT. From top to bottom: (a) the magnetic field components in \correction{the} RTN reference frame and the magnetic field modulus, (b) proton velocity \correction{radial component $V_R$},  (c) angle between the magnetic field and the \correction{radial direction $\Theta_{\mathbf{BR}}$}, 
 (d) electron density $N_e$, (e) proton temperature $T_p$ (in blue) and core electron temperatures $T_{e,core}$ (in black), (f) core electron and proton plasma beta, (g) ion characteristic scales, $d_i$ and $\rho_i$. \label{fig:plasmadata} }
\end{figure}

We analyze \correction{a 5~hour} time interval during the first perihelion, on November 6, 2018, [00:00, 05:00]~UT, when the spacecraft at the distance of 0.17~au from the Sun measured the solar wind emerging from the small equatorial coronal hole  \citep{2019Bale,2019Kasper}.
The magnetic field during the chosen time interval is particularly highly-disturbed due to the presence of high-amplitude structures \cite[including switchbacks,][]{2019Bale,2020Perrone}. 
The duration of the chosen interval is long enough to resolve the inertial range of MHD turbulence, but not too long, so that the PSP is magnetically connected to the same coronal hole and the PSP position is nearly at the same  radial distance from the Sun. 

We use the merged magnetic field measurements of two magnetometers:  FIELDS/Fluxgate Magnetometer and Search Coil \citep{Bale2016,2020Bowen}. 
These data have a $3.4$~ms time resolution, which allows us to resolve a wide range of scales, from \correction{the} MHD inertial range to \correction{the} sub-ion range. Due to the \correction{anomaly of the Search Coil at one of its} \correction{axes} (\correction{which} happened \correction{in} March 2019), the full merged vector of the magnetic field is accessible only for the first perihelion \citep{2020Bowen}.
Figure~\ref{fig:plasmadata}(a) shows the magnetic field magnitude $B(t)$ in black and three components in \correction{the} RTN coordinate frame in color. 
The magnetic field vector fluctuates around $\langle \mathbf{B} \rangle=(-47,63,-5)$~nT. Its magnitude is nearly constant $|\mathbf{B}|=98\pm5$~nT. The angle between the magnetic field and \correction{the radial direction changes from $25^{\circ}$ to $180^{\circ}$ as shown in Figure~\ref{fig:plasmadata}(c), with a dominance around quasi-perpendicular orientation, $\langle \Theta_{\mathbf{B}\mathbf{R}}\rangle = 121^{\circ}$ with standard deviation $\sigma( \Theta_{\mathbf{B}\mathbf{R}}) = 26^{\circ}$. } 

To characterize plasma \correction{bulk velocity,} we use \correction{the} SWEAP/SPC Faraday cup instrument \citep{Kasper2016}.
Proton velocity $\mathbf{V}$ \correction{is} estimated from the 1st moment of the distribution function. 
\correction{Figure~\ref{fig:plasmadata}(b) shows its radial component  $V_{R}$. } 
The mean  velocity is $\langle \mathbf{V}\rangle=(330,70,8)$~km/s, \correction{making a non-zero} angle with the radial direction: $\langle\Theta_{\mathbf{VR}}\rangle=14^{\circ}$. \correction{This small deviation from the radial of the solar wind speed  might not be accurate due to the field of view of the instrument together with the high angular satellite velocity during the considered time interval [Michael Stevens, private communication]. It may introduces a relative error of about 3\% on the estimates of spatial scales $\ell$ using Taylor hypothesis.} 

We use RFS/FIELDS quasi-thermal noise (QTN) electron plasma data to characterize electron plasma parameters  \citep{2020Moncuquet}. Electron density $N_{e}$ is determined from the electrostatic fluctuations at the electron plasma frequency, and it is shown in Figure~\ref{fig:plasmadata}(d). 

Proton temperature $T_{p}$ is estimated from the second moment of the distribution function measured by \correction{the} SWEAP/SPC instrument. 
Figure~\ref{fig:plasmadata}(e) shows 
QTN electron core temperature $T_{e,core}$ \correction{(in black)} and proton temperature $T_{p}$ \correction{(in blue). The mean temperatures are $\langle T_{e,core}\rangle = 28.5$~eV and $\langle T_p\rangle = 15$~eV}.

We consider solar wind fluctuations over a wide range of timescales. The minimum time scale is determined by the temporal resolution of the magnetic data $dt=3.4\cdot 10^{-3}$~s, and the maximum by the total duration of the time interval $T=5$~h.
\correction{The spatial scales of the corresponding range,} estimated using \correction{the} Taylor hypothesis with $V= 340$~km/s, $\ell \in [1,10^5]$~km are much larger than the Debye length $\lambda_D \simeq 2$~m, then the plasma is quasi-neutral. 
During the \correction{analyzed} time period, alpha particle abundance  \correction{has not been measured, but it can be estimated as}
$A_{\alpha}=N_{\alpha}/N_{p}<5\%$ 
\citep{2007Kasper,2019Alterman,Mingzhe2021}. 
The quasi-thermal noise spectroscopy provides \correction{a} more accurate measurement of the density than particle detectors, so we use $N_{p}=N_{e}$ and calculate proton plasma beta $\beta_{p}$ using the electron density: $\beta_p= N_e k T_p/(B^2/2\mu_0)$, with $\mu_0$ being the magnetic permeability. Plasma beta for core electrons is defined as  
$\beta_{e,core}=N_e k T_{e,core}/(B^2/2\mu_0)$. Both plasma $\beta$ parameters are well below unity as shown in Figure~\ref{fig:plasmadata}(f), \correction{$\langle \beta_{p}\rangle=0.22$, $\langle \beta_{e,core}\rangle=0.44$}. 
Figure~\ref{fig:plasmadata}~(g) shows \correction{the} time variation of the ion characteristic spatial scales $d_i=c/\omega_{pi}$, where $\omega_{pi}=\sqrt{4\pi N_p e^{2}/m_p}$ and $\rho_i=V_{th,\perp}/\omega_{ci}$, where \correction{ $V_{th,\perp}=\sqrt{2k_{B}T_{\perp}/m_p}$} is the perpendicular proton thermal speed, $\omega_{ci}=eB/m_{p}c$ is the proton gyrofrequency.

\section{MHD, ion, and sub-ion \correction{range} identification} 
\label{sec:spectrum}

\begin{figure}[ht!]
	\centering
	\includegraphics[width=1.04\linewidth]{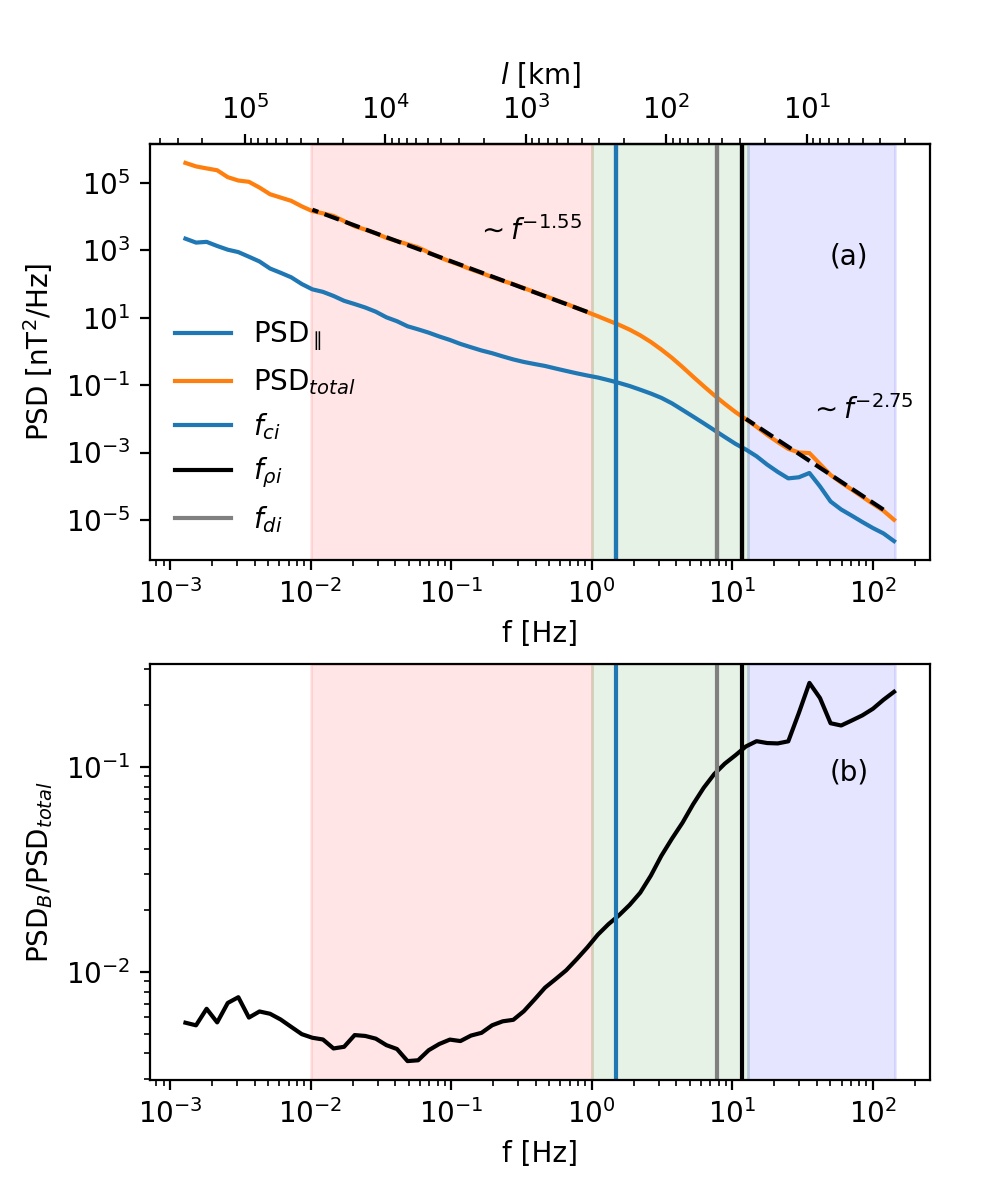}
	\caption{
(a) Magnetic field total spectrum $S_{total}$ in orange and magnetic field modulus spectrum, \correction{that is the proxi of} the parallel fluctuations spectrum $S_{\parallel}$  (Equation (\ref{eq:S_parallel})), and (b) the ratio $S_{\parallel}/S_{total}$. The vertical lines show the characteristic ion scales: ion cyclotron frequency $f_{ci} $ (in blue), and the frequencies computed with the Doppler shifted ion gyroradius $f_{\rho i}$ (in black) and the Doppler shifted ion inertial length $f_{di}$ (in grey). The frequency ranges are highlighted: MHD in red, ion scales in green and sub-ion scales in blue. \label{fig:spectral}}
\end{figure}

First\correction{,} we describe the magnetic field spectral properties of the \correction{analyzed} time interval. 
We apply wavelet transform with Morlet mother function \citep{1998Torrence}:
  \begin{equation}  \label{eq:psi_0}
  \psi_{0}(t)=\pi^{-1/4} e^{-t^{2}/2} e^{i\omega_{0}\, t},
  \end{equation}
where $\omega_{0}=6$  is the angular frequency of oscillations in the mother function (with normalized time).
The wavelet transform of the magnetic field component $B_{i}(t)$ is defined as the convolution of $B_{i}(t)$ with scaled, translated, and normalized $\psi_{0}(t)$  to have mother function $\psi$ with unit energy:
  \begin{equation}  \label{eq:W[B_i]}
   W[B_{i}](t,\tau)=\sum\limits_{n'=0}^{N-1}B_{i}(t') \psi^{\ast} [ (t'-t)/\tau ]
   \end{equation}
where the sign $^{\ast}$ indicates \correction{the} complex conjugate.

Wavelet coefficients are influenced by the edge effects. \correction{The} cone of influence (COI) curve separates the region of scales where edge effects become important as a function of time. To avoid this edge effect we consider a maximum scale equal to $\tau_{max}=10^3$~s. The intercept of $\tau_{max}$ with \correction{the} COI curve determines the time sub-interval $T'=$ [00:22:49, 04:37:11]~UT, where wavelet coefficients at the scales $\tau<\tau_{max}$ are not
influenced by the edge effect. 

Figure~\ref{fig:spectral}(a), orange line, shows the total magnetic field power spectral density (PSD)  $S_{total}(\tau)$, calculated using the time-averaging over the subinterval $T'$ :
   \begin{equation} \label{eq:S_total}
   S_{total}(\tau)=\frac{2 \delta t^{2}}{T'} \sum\limits_{t\in T'} \sum\limits_{i=R,T,N} |W[B_{i}](t,\tau)|^2,
   \end{equation}
where $\delta t=0.008$~s is the time-step of the PSP merged magnetic field data. The relation between Fourier frequencies $f$ and time scales $\tau$ is $f\simeq 1/\tau$ for the Morlet wavelets with $\omega_0=6$. \correction{In} Figure~\ref{fig:spectral}(a), the blue line shows the PSD of compressive magnetic fluctuations. 
 Compressive fluctuations are approximated here by the variation of magnetic field modulus.
 Indeed, this approximation is valid
 if the level of the fluctuations is significantly lower than the mean field $B_0$, i.e., $\delta B/B_0 \ll 1$ \citep{2016Perrone}:
  \BE   \label{eq:delta_B2}
   \delta (|B|^2)=|\vB_0 + \delta \vB|^2-|\vB_{0}|^2 \approx 2 \delta B_{\parallel}B_{0} \approx \delta (B_{\parallel}^2)
   \EE
In the inertial range and at higher frequencies the condition $\delta B/B_0\ll 1$ is valid.
So we calculate the parallel PSD, $S_{\parallel}(\tau)$, as was done \correction{by} \cite{2016Perrone}:
  \begin{equation}    \label{eq:S_parallel}
  S_{\parallel}(\tau)=\frac{2 \delta t^{2}}{T'} \sum\limits_{t\in T'} |W[{|B|}](t,\tau)|^{2}
  \end{equation}
       
As we can see from Figure~\ref{fig:spectral}(a), 
$S_{total}(f)\sim f^{-1.55}$ within the inertial range $10^{-2}< f < 1 $~Hz,  
in agreement with \cite{2020Chen}. Approaching ion kinetic scales, the spectrum steepens. 
The ion transition range, or simply ion scales, is present where the spectrum changes continuously its slope \citep[][]{2013Alexandrova,Kiyani2015}. It is observed here nearly between the ion cyclotron frequency $f_{ci}=eB/2 \pi m_i=1.4$ Hz and the frequency of the Doppler-shifted ion gyroradius $f_{\rho i}=V/2\pi \rho_i=11.4$ Hz. 
The frequency of the Doppler-shifted ion inertial length $f_{di}=V/2\pi d_i$ is in between these two frequencies.
At $f > 13$~Hz (sub-ion scales), the spectral index stabilizes at $-2.75$, in agreement with what is observed at 0.3 and 1~au between ion and electron scales \citep{Alexandrova2009,Chen2010,Alexandrova2012,Alexandrova2021}.

Based on the magnetic field spectral properties and characteristic plasma scales ($f_{ci}$, $f_{\rho i}$ and $f_{di}$)
we define the following frequency ranges $\Delta f_j$, 
shown as transparent color bands in Figure~\ref{fig:spectral}: 
\begin{equation}
\label{eq:frequency_ranges}
   \Delta f_j = 
    \begin{cases}
    (10^{-2}, 1) \text{ Hz} & \text{MHD inertial range} \;\text{(in red)}\\
    (1 , 13)     \text{ Hz} & \text{ion scales} \;\text{(in green)}\\
    (13, 128)    \text{ Hz}& \text{sub-ion range} \;\text{(in blue)}  
    \end{cases}       
\end{equation}

The corresponding timescale ranges 
$\tau_{j}$ will be used later in this article, and the index $j$ here and further in the article refers to one the following ranges:
\begin{equation}
\label{eq:timescale_ranges}
   \tau_{j} :
    \begin{cases}
    \tau_{MHD}= (1,100)        \text{ s} & \\
    \tau_{ion}=(0.08,1)       \text{ s} & \\
    \tau_{subion}=(0.008, 0.08) \text{ s} & 
    \end{cases}       
\end{equation}

The ratio of compressible fluctuations to the total power spectral density $S_{\parallel}/S_{total}$ is shown in Figure~\ref{fig:spectral}(b).  In the inertial range, parallel magnetic fluctuations are much less energetic than perpendicular ones ($\delta B_{\|} \ll \delta B_{\perp}$), as is usually observed in the solar wind. 
At the sub-ion scales, the fraction of the parallel $S_{\parallel}(\tau)/S_{total}(\tau)$  increases up to $\simeq 0.2$, which is consistent with the results of \citet{2012Salem} at 1~au. The authors suggested that the observed spectral ratio can be explained by
the presence of the kinetic Alfvén wave (KAW) cascade with  nearly perpendicular wavevectors ($k_{\perp}\gg k_{\|}$). 
However, analyzing Cluster measurements \citep{2017Lacombe} and 2D hybrid numerical simulation \citep{2020Matteini} found that \correction{the} asymptotic compressibility value at sub-ion scales doesn't match perfectly the KAW prediction. 
Finally, recent numerical simulations indicate that coherent structures, rather than waves, are energetically dominant on sub-ion scales \citep{2021Papini}.

\section{Detection of coherent structures from MHD to sub-ion scales} \label{sec:detection}
In this section\correction{,} we describe the methodology to detect the structures from MHD down to sub-ion scales.

\subsection{Local intermittency measure}
\label{sec:LIM}
\begin{figure}[t!]
	\centering
	\includegraphics[width=1.0\linewidth]{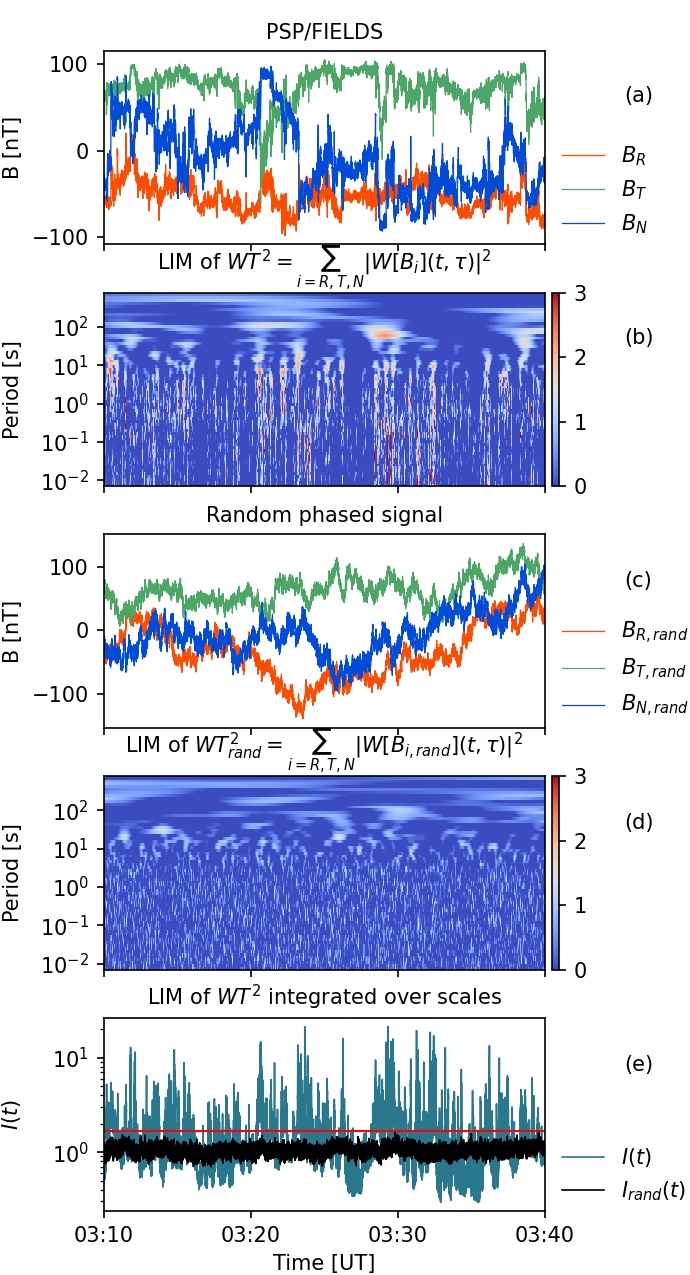}
\caption{ 
A 30 minutes zoom, [03:10,03:40]~UT, within the analyzed time interval of 5 hours on November 6, 2018.
From top to bottom: 
  (a)~magnetic field in \correction{the} RTN reference frame, 
  (b)~LIM of the magnetic fluctuations of the total energy $L(t,\tau)$, Equation~(\ref{eq:L}),
  (c)~artificial magnetic field $\mathbf{B}_{rand}$ with random phases and the same Fourier amplitudes as original magnetic field measurements, 
  (d)~LIM of the artificial signal $L_{rand}(t,\tau)$, 
  (e)~the comparison of the integrated LIMs $I(t)=\langle L(t,\tau) \rangle_{\tau \in [10^{-2},10^{3}]~s}$ (blue), and the $I_{rand}(t)=\langle L_{rand} (t,\tau) \rangle_{\tau \in [10^{-2},10^{3}]~s}$ (black). 
The horizontal red line shows $I_{threshold}=\text{max}(I_{rand}(t))$  as defined in Figure \ref{fig:intermittency_hist}.
}
\label{fig:intermittency}
\end{figure}

\begin{figure}[ht!]
	\centering
	\includegraphics[width=1.05\linewidth]{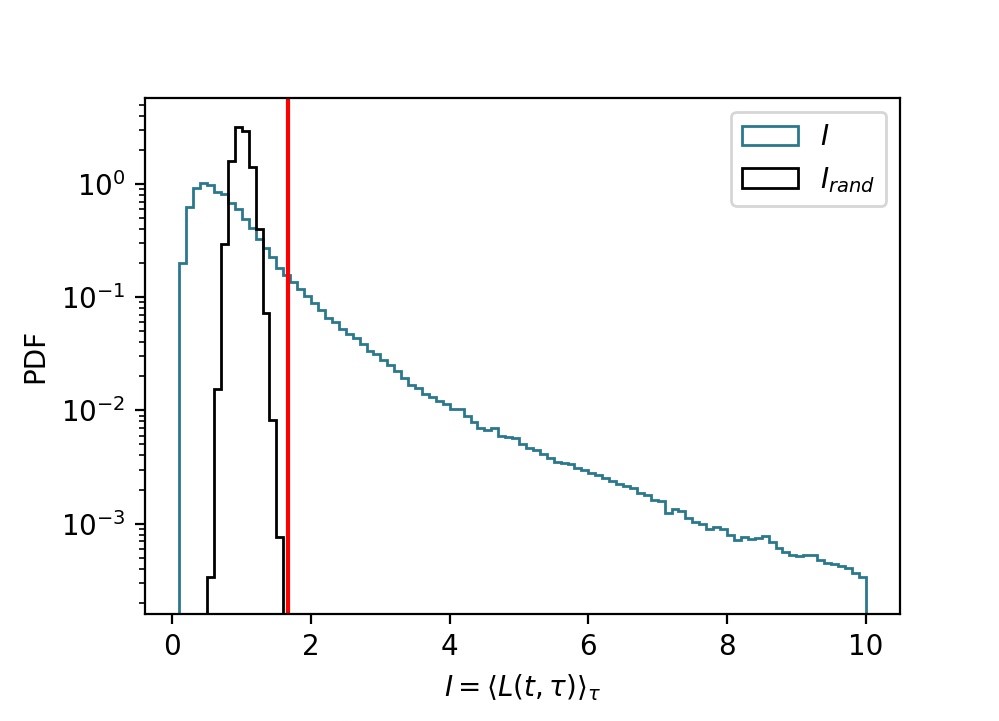}
\caption{Histograms of the integrated LIM $I(t)$ and the random phased integrated LIM $I_{rand}(t)$.
The threshold $I_{threshold}=\text{max}(I_{rand}(t))$ is shown by the red vertical line.}
\label{fig:intermittency_hist}
\end{figure}

We use the Local Intermittency Measure (LIM) $L(t,\tau)$ \citep{Farge1992} based on Morlet wavelets to detect the structures. 
  The value $L(t,\tau)$ shows the total energy of fluctuations at a given moment in time $t$ at a given time scale $\tau$, relative to the average energy at that scale:
   \begin{equation}   \label{eq:L}
    L(t,\tau)=\frac{\sum_{i=R,T,N}|W[B_{i}](t,\tau)|^{2}}{\langle \sum_{i=R,T,N}|W[B_{i}](t,\tau)|^{2} \rangle_{t\in T'}}
    \end{equation} 
where $T'$ is the analyzed time interval.

In Figure~\ref{fig:intermittency} we show a \correction{30-minute} zoom within $T'$.
Panel~(a) gives RTN components of the measured $\mathbf{B}$. Panel (b) shows the observed $L(t,\tau)$.
The vertical elongations of enhanced $L(t,\tau)$ values are due to coupled (or coherent) phases of the fluctuations \citep{2016Lion,2016Perrone,Alexandrova2020_hdr}.
Indeed, to see this point better, we construct an artificial signal that has the same Fourier spectrum as the original magnetic field measurements, but with random phases \citep{Hada2003,Koga2003}.  
This synthetic signal $\mathbf{B}_{rand}$ is shown in Figure~\ref{fig:intermittency}(c), while the corresponding LIM $L_{rand}(t,\tau)$ is shown in the panel~(d). The energy distribution of the synthetic signal is incoherent (randomly distributed in the $(t,\tau$)--plane), i.e., peaks of $L_{rand}(t,\tau)$ at different $\tau$ are not observed at the same time.  
Therefore, the vertical elongations in the observed $L(t,\tau)$ correspond to magnetic fluctuations with coupled phases across scales where the elongation is observed. The high energy of these events with respect to the mean is a sign of 
intense coherent structures formed in the turbulent medium \citep[e.g.,][]{Farge1992, 2019Bruno}.
So, we observe coherent structures which extend from inertial to sub-ion timescales. Using the Taylor hypothesis, the timescale range $\tau \in \tau_{all} = [10^{-2},10^{3}]$~s can be converted into the spatial range $\ell=V\cdot \tau \in [3, 3 \cdot 10^{5}]$~km, \oa{or $\ell/\rho_i=6[0.1,10^4]$ in terms of the ion Larmor radius ($\rho_i=5$~km). }

The difference between random-phased signal and original magnetic field data suggests a methodology for detecting the central times of coherent structures. \oa{Specifically, we sum the 
LIM over the timescale range $\tau_{all}=[10^{-2},10^{3}]$~s, where wavelet timescales $\tau$ are logarithmically spaced with the base~2, see Eq.~(9) in \citep{1998Torrence}:} 
\BE \label{eq:I}
I(t)=\sum_{\tau \in \tau_{all}} L(t,\tau)
\EE

Figure~\ref{fig:intermittency}(e) shows $I(t)$ (blue line), random phased integrated LIM $I_{rand}(t)$ (black line) and the threshold $I_{threshold}=\text{max}(I_{rand}(t))$ (red horizontal line). 
The local maxima of $I(t)>I_{threshold}$ \correction{gives} the central times of the coherent structures present in the original signal. 
\correction{In the following,} we refer to this method as \textit{the integrated LIM selection}.

The comparison of original $I(t)$ and random phased $I_{rand}(t)$ distributions is shown \correction{in} Figure~\ref{fig:intermittency_hist}. The $I_{rand}$ distribution (in black) is close to Gaussian with a mean of 1 (because of the normalization and random phases).
On the contrary, $I(t)$ (in blue-azure) has a long tail of extreme 
values due to the presence of coherent structures integrated over all time scales.

The integrated LIM selection does not have a predetermined scale at which the structure is searched for but it is preferentially focused on scales where the vertical enhancements in the LIM $L(t,\tau)$ are observed. 
Applying it on $T'=$[00:22:49,04:37:11] on 6 November 2018, we find $N=9485$ structures. 
If we define the filling factor of the structures as the normalized total time duration where the integrated LIM is over the threshold:
\begin{equation}\label{eq:filling_factor}
  P=\text{Time}(I(t)>I_{threshold})/T',  
\end{equation}
we find that the structures cover $14\%$ of the analyzed time interval $T'$.

In this paper, we will also use the integrated LIM over the reduced time-scale ranges, to understand in more detail the nature of the structures at  MHD, ion, and subion scales, where physics is different. So, we can define integrated LIM $I_{i}=(I_{MHD},I_{ion},I_{subion})$ over the corresponding range of timescales $\tau_{j}=(\tau_{MHD},\tau_{ion},\tau_{subion})$, defined in Equation~(\ref{eq:frequency_ranges}):
\BE \label{eq:I_j}
I_j(t)=\sum_{\tau \in \tau_{j}} L(t,\tau)
\EE
Similarly, integrating $L_{rand}(t,\tau)$ over $\tau_{j}$ we define random phased integrated LIM $I_{rand,j}(t)$. 
Thus, we can find the central times of the structures within these 
scale-bands as the times of the local maxima for $I_j(t)>I_{threshold,j}=\text{max}(I_{rand,j}(t))$.

This band-integrated LIM selection allows us to see how the number of the structures and filling factor changes with \correction{the} band \correction{of scales}. 

\oa{In order to count isolated coherent structures, we find continuous time intervals when $I_j(t)>I_{threshold,j}$}. $N_{j}$ denotes the number of isolated events at each range of scales.
We define the filling factor $P_j$ as follows
   $$P_{j}=\text{Time}(I_j(t)>I_{threshold,j}) \,/\, T'.$$
We find a relatively small number of MHD scale structures ($N_{MHD}=$196) with high filling factor ($P_{MHD}=$12\%), compared to $P_{ion}=$7\% and $P_{sub-ion}=$6\% for much more numerous ion scale structures ($N_{ion}=2028$) and sub-ion scale structures ($N_{sub-ion}=11167$).
We remark, that our estimations of $P$ are conservative, as far as only time where LIM is over the threshold is counted, but the structure's field decreasing from its center \correction{exists} outside of the time where the energy of the structure is concentrated, e.g., \citep{2016Perrone}. 
So, the filling factor can be more than twice \correction{as large as indicated herein}.  
Finally, numerous small-scale events populate larger ones and may exist outside them as well.

\subsection{Magnetic field at different 
scales} \label{sec:Filtering}
\begin{figure}[ht!]
	\centering
	\includegraphics[width=1.05\linewidth]{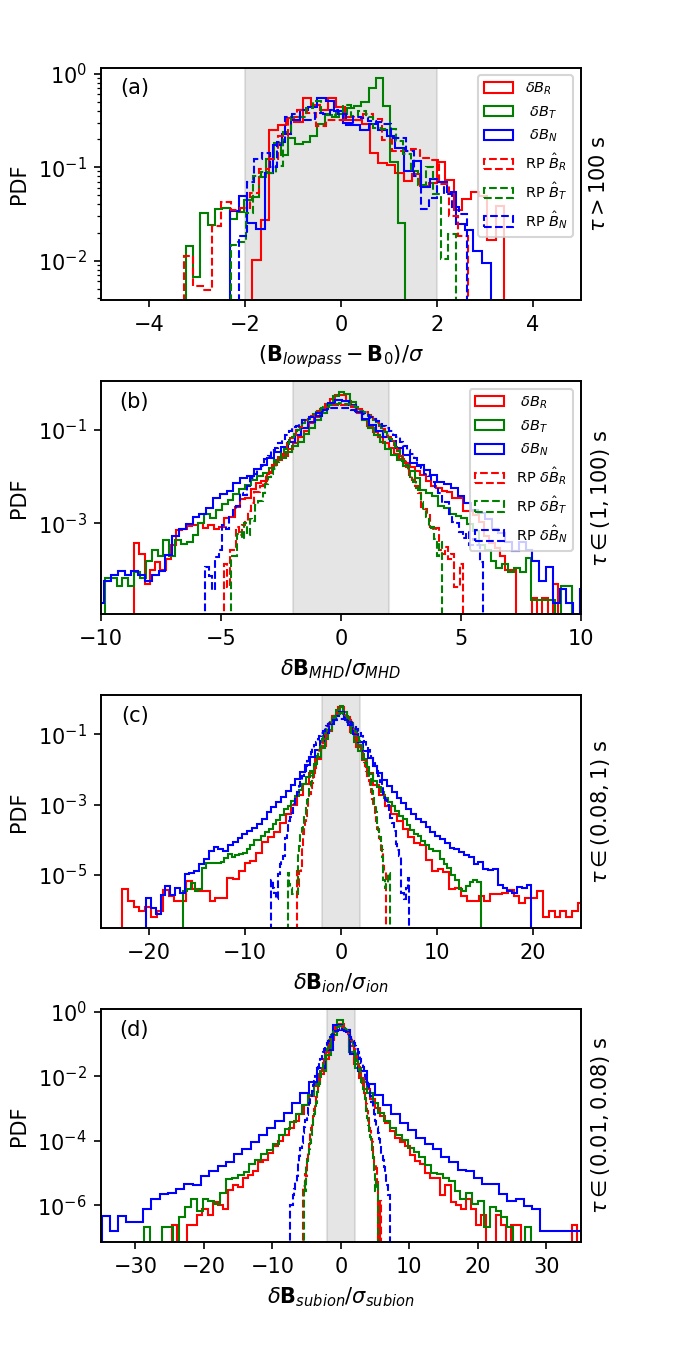}
\caption{Histograms of magnetic field fluctuations (solid) compared to the signal with random phases (RP, dashed). The first panel from the top shows the centered lowpass-filtered fluctuations of the magnetic field. Panels (b-d) show bandpass-filtered fluctuations on MHD inertial, ion kinetic, and sub-ion scales, respectively. The horizontal axis is normalized to the standard deviation of the random-phased signal. The area within two standard deviations of the random-phased signal is highlighted in gray.
}
\label{fig:filtering_intermittency}
\end{figure}

\begin{table*}
\centering
\begin{tabular}{lllllllllllll}  
& \multicolumn{1}{c}{$\sigma_{R}$ } &\multicolumn{1}{c}{$\sigma_{T}$}& \multicolumn{1}{c}{$\sigma_{N}$ }  &\multicolumn{1}{c}{$\sigma_{noise}$ }  & \multicolumn{1}{c}{$s_R$} & \multicolumn{1}{c}{$s_T$} & \multicolumn{1}{c}{$s_N$} & \multicolumn{1}{c}{$\langle |s| \rangle $} & \multicolumn{1}{c}{$\kappa_R$}& \multicolumn{1}{c}{$\kappa_T$}& \multicolumn{1}{c}{$\kappa_N$}& \multicolumn{1}{c}{$\langle \kappa \rangle$}\\
\hline
\hline
Low-pass   & 37   & 31   & 35   & -  & 1.1  & -1.1  & 0.5   &0.9&  3.8  & 3.8  & 3.1  & 3.6\\
MHD        & 9    & 8.6  & 11   &7.8 & 0.2  & -0.5  & 0.01  &0.2&  7.7  & 8.7  & 6.5  & 7.6\\
Ion scales & 1.5  & 1.6  & 2.2  &1.55& 0.1  & -0.2  & -0.01 &0.1&  16.3 & 14.0 & 10.8 & 13\\
Sub-ion    & 0.12 & 0.12 & 0.15 &0.11& 0.01 & -0.01 & -0.04 &0.02& 11.6 & 19.3 & 24.4 & 18\\
\end{tabular}
\caption{Main parameters of the distributions shown in Figure \ref{fig:intermittency_hist} for the magnetic field components in \correction{the} RTN coordinates. From left to right: standard deviation $\sigma_{RTN}$ [nT],  average standard deviation $\sigma_{noise}$ [nT] (in absence of coherent structures), skewness $s_{RTN}$, average absolute skewness $\langle |s| \rangle$, kurtosis $\kappa_{RTN}$, and the average kurtosis $\langle \kappa \rangle$. }
\label{table:0}
\end{table*}

Thanks to Morlet wavelets and LIM we know now the central times of the structures covering all scales and the ones within different scale bands. 
In order to study magnetic field fluctuations $\delta\mathbf{B}$ in the physical space around these central times, within different scale bands, we use \correction{a} band-pass filter for fluctuations on frequency ranges given by Equation~(\ref{eq:frequency_ranges}) and shown by color bands in Figure~\ref{fig:spectral}. 
We complete this analysis by studying the large scale fluctuations of $\mathbf{B}_{lowpass}-\mathbf{B}_{0}$ where the mean field $\mathbf{B}_{0}$ is defined as the average field over the time interval $T'$.
We use finite impulse response (FIR) Hamming low-pass filter with a cut-off frequency of $10^{-2}$~Hz to calculate the large scale magnetic field fluctuations of $\delta\mathbf{B}=\mathbf{B}_{lowpass}-\mathbf{B}_{0}$ \citep{Smith1997}.

Figure~\ref{fig:filtering_intermittency} shows distributions of the filtered magnetic field (solid lines) compared to the filtered signal with random phases (dashed lines). Panel~(a) shows the lowpass-filtered fluctuations of the magnetic field. Panels~(b-d) show the FIR Hamming bandpass-filtered fluctuations on MHD, ion, and sub-ion scales, respectively.
At each band of scales, we characterize the amplitude of incoherent fluctuations as follows:
\begin{equation}\label{eq:sigma_noise}
\sigma_{noise,j}=\text{std}(\delta B_j(t \in T_{\text{no struct}}))
\end{equation}
where $T_{\text{no struct}}=\text{Time}(I_j(t)<I_{threshold,j})$.
The gray area in panels (b-d) is bounded by $\delta \mathbf{B}_{rand,j} /\sigma_{noise,j}=\pm 2$.

The random phase signal fluctuations have Gaussian distributions at all scales (Figure~\ref{fig:filtering_intermittency}). The observed $\delta\mathbf{B}$ show scale-dependent deviation from Gaussianity. Table \ref{table:0} gives the moments of the observed distributions for 3 components at 4 different scale ranges.  The distributions have non-zero skewness $s$ (a normalized measure of a distribution asymmetry). 
The fourth normalized moment, kurtosis $\kappa$, increases from 3-4 at large scales, up to 12-24 at sub-ion scales. In comparison, Gaussian noise has $s=0$ and $\kappa=3$. 

Distributions of lowpass magnetic field components 
are asymmetric with respect to zero, especially radial and tangential (see Figure \ref{fig:filtering_intermittency}(a)). The skewness of those components has high absolute value and opposite signs: $s_R=1.1$ and $s_T=-1.1$.
The lowpass magnetic field distributions \correction{don't} have pronounced non-gaussian tails, so the kurtosis $\kappa$ \correction{is} slightly above 3, so close to the gaussian noise value (see Table \ref{table:0})).

In the inertial and smaller scale ranges, the distributions have weaker asymmetry ($\langle |s| \rangle=(|s_R|+|s_T|+|s_N|)/3 \leq 0.2$).  
Non-gaussian tails are identified at MHD scales (Figure \ref{fig:filtering_intermittency}(b)) and become even more pronounced at ion and sub-ion scales (Figure \ref{fig:filtering_intermittency}(c-d)). 
The kurtosis $\kappa_T$ and $\kappa_N$ monotonically increase from MHD to subion scales (see Table \ref{table:0}). The kurtosis of the radial magnetic field $\kappa_R$ component is growing from MHD to ion scales and then \correction{decreases} at subion scales. 
This behavior of $\kappa_R$ can be explained by the proximity of the SCM noise, which starts to influence $\delta B_{R,subion}$, the weakest of the 3 components of magnetic fluctuations at these scales, see the red PDF in Figure \ref{fig:filtering_intermittency}(d).

\section{Model structures} \label{sec:models}
\begin{figure*}[ht!]
	\centering
	\includegraphics[width=1.05\linewidth]
{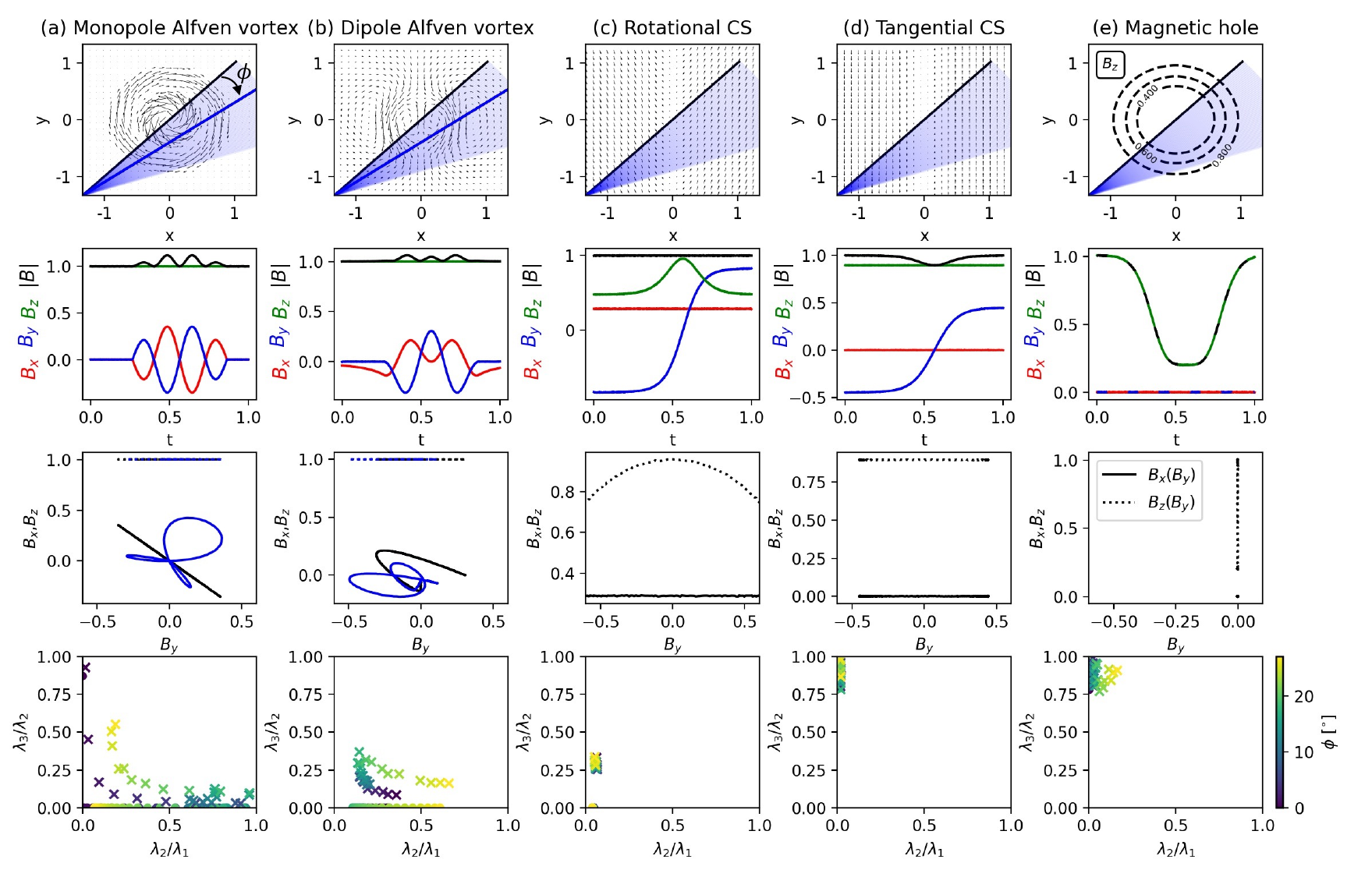}
  \caption{ 
Simulation of the spacecraft crossing (a) a monopole Alfvén vortex, (b) a dipole vortex, (c) a rotational and (d) a tangential discontinuities, and (e) a magnetic hole. The first row shows the magnetic field vector in the plane perpendicular to the background magnetic field. 
The sector, shown in blue, is a set of trajectories crossing the structure at different angles in order to collect statistics of MVA eigenvalues. The panels in the second line show the magnetic field in the MVA frame of reference, which would be measured by the spacecraft when it crosses the structure along the black trajectory. Panels in the third line show the hodograph - indicating polarisation for off-center (blue) and central (black) trajectories. 
  The bottom row shows the eigenvalue ratios for the set of trajectories shown within the blue cone in the first row in the presence of noise, with $\epsilon=0.001$ (circles) and $\epsilon=0.1$ (crosses), where $\epsilon$ is defined in Equation~\ref{eq:epsilon_struct}. The trajectory angle $\phi$, defined in the top left panel, is coded with colors (see the color scale at the right bottom). 
\label{fig:models}}
\end{figure*}

In this Section\correction{,} we discuss several models of the coherent structures.
This gives us a necessary background to determine the dominant type of structures in the large statistics of events. 
\oa{These models have been developed in the MHD framework. Therefore they are not applicable \textit{a priori} for ion and subion scales. But kinetic-scale turbulence may be described with fluid-like equations, which are structurally similar to reduced MHD equations.
Therefore, it is reasonable to expect that similar types of structures (vortices, current sheets) can be distinguished among coherent structures on ion and subion scales.}
This is why we compare these models of coherent structures at kinetic scales as well.

The trajectory of a spacecraft across a structure  matters for the polarization and the amplitude anisotropy of magnetic fluctuations. That is why we will explore the polarization and the \oa{Minimum Variance Analysis \citep[MVA,][]{MVA} results} as a function of the spacecraft trajectory across the model structures. 

\subsection{Alfvén vortices}
Alfvén vortices are cylindrically symmetric coherent structures that were introduced by \cite{1992Petviashvili}. Within these vortices, the generalized Alfv\'en relation $\delta\mathbf{V}_{\perp}/
V_A=\xi \,\delta\mathbf{B}_{\perp}/B_0$ is verified, with $\xi$, which may be different from 1.

\subsubsection{Monopole Alfvén vortex}\label{sec:Monopoles}
Figure~\ref{fig:models} (column (a), top row) shows the crossing of a monopole vortex \citep{1992Petviashvili}, details of the model are described in appendix~\ref{sec:Alfvén_monopoles}.
The set of trajectories, selected to  cross the vortex, is shown by the blue transparent cone on the top left panel of Figure~\ref{fig:models}. The set is \correction{parameterized} by the angle $\phi$. Two trajectories, central ($\phi=0^\circ$) and off-center ($\phi=10^\circ$), are shown in black and blue lines correspondingly.
The second panel from the top shows the three magnetic field components of the monopolar vortex crossed by a spacecraft along the black trajectory in the top panel. 
The third panel shows the dependencies $B_x(B_y)$ and $B_z(B_y)$ for both central and off-center trajectories (black and blue lines respectively).  The off-center trajectory has 'clover'-like polarisation in $B_x(B_y)$ (blue curve). In \correction{the} case of the crossing through the center, the polarisation is linear (black line). 

Figure~\ref{fig:models} (column (a), bottom row) shows the MVA eigenvalue ratio $\lambda_{3}/\lambda_{2}$ as a function of $\lambda_{2}/\lambda_{1}$ for 50 different trajectories (see the blue cone in the top panel). The eigenvalues are ordered as $\lambda_1 \geq \lambda_2 \geq \lambda_3$, with the eigenvector $\mathbf{e_3}$ being the minimum variance direction.
The color between violet and yellow indicates the angle $\phi$ of the trajectories: $\phi=0^{\circ}$ corresponds to the crossing through the center and $\phi=25^{\circ}$, to the side-crossing. 
In this plot\correction{,} we test the effect of an added noise with a relative amplitude $\epsilon$ defined by
   \begin{equation}    \label{eq:noise}
   \epsilon=\delta B_{noise}/\delta B_{\perp}
   \end{equation} 
where $\delta B_{noise}$ is the noise amplitude and $\delta B_{\perp}$ is the amplitude of the vortex.
The eigenvalue ratios $\lambda_{2}/\lambda_{1}$ and $\lambda_{3}/\lambda_{2}$ are dependent on $\epsilon$. 
The results for two levels of noise are shown: $\epsilon_{1}=0.001$ with filled circles, and $\epsilon_{2}=0.1$ with crosses.
In case of negligible noise, the points are located along the x-axis \correction{(see circles)}.  
For larger $\epsilon$, the eigenvalues become more comparable \correction{and the plane $(\lambda_{2}/\lambda_{1},\lambda_{3}/\lambda_{2})$ fills-up: the data form the \textit{shape of a croissant} (see crosses). In case of a very large noise (and in case of random fluctuations), the eigenvalues become of the same order, i.e., $\lambda_{2}/\lambda_{1} \sim \lambda_{3}/\lambda_{2} \sim 1 $ (not shown). }
In Section~\ref{sec:minvar}, $\epsilon$ is estimated from observations.

For the majority of trajectories, except central ($\phi<3^{\circ}$) and extreme-off-center/tangential ones ($\phi>22^{\circ}$), the minimum variance direction $\mathbf{e}_3$ is well-defined ($\lambda_{3}/\lambda_{2} \sim 0$) and it is parallel to the axis of the vortex. Indeed, the vortex model describes $\delta B_{\perp}$ and assumes $\delta B_{z}=0$. So, in observations, $\mathbf{e}_3$ (when it is well-defined) is a good approximation for the vortex axis. 
As far as the vortex cylinder is field-aligned, the angle between $\mathbf{e}_3$ and $\mathbf{B}_0$ must be small, $\theta_{B_{0},3} \sim 0^{\circ}$.

In case of the central crossing ($\phi<3^{\circ}$), only $\mathbf{e}_1$ is well-defined, because $\lambda_2/\lambda_1 \sim 0$, $\lambda_3/\lambda_2 \sim 1$. 
In this case the eigenvector of maximal variance $\mathbf{e}_1$ is perpendicular to the crossing trajectory ($\mathbf{e}_1\perp \mathbf{V}$) and to the background magnetic field ($\mathbf{e}_1\perp \mathbf{B}_0$). Therefore, $\theta_{V,1} \sim 90^{\circ}$ and $\theta_{B,1} \sim 90^{\circ}$ are expected in observations.

In addition, for the vortex to be observable, the spacecraft must cross it along a trajectory inclined at a sufficient angle relative to the vortex axis, so $\theta_{BV} \ne 0^{\circ}$ and $\theta_{V,3} \ne 0^{\circ}$ (if $\mathbf{e}_3$ is well-defined).

\subsubsection{Dipole Alfvén vortex}\label{sec:Dipoles}

In Figure~\ref{fig:models} column (b), \correction{the} top panel shows the magnetic field of the dipole vortex. This type of Alfv\'en vortices is particular, because \correction{its} axis is inclined with respect to the background magnetic field and it is propagating, see details of the model in appendix~\ref{sec:Alfvén_dipoles}. 
The magnetic field components are symmetric in time around the vortex axis
(while for the monopole vortex they are anti-symmetric).
The magnetic polarisation (third panel from the top) is different for the crossing at the vortex center (black trajectory), and the side crossing (blue trajectory). 

Figure~\ref{fig:models}~(column (b), bottom panel) gives the minimal variance eigenvalues ratios for two noise levels. 
In case of the low noise, $\epsilon=0.001$, $\lambda_{3}/\lambda_{2} \sim 0$ and $\lambda_{2}/\lambda_{1} \in [0.1,0.6]$ (filled circles). 
For $\epsilon=0.1$, as for the monopole vortex, both ratios increase: the points in the $(\lambda_{2}/\lambda_{1},\lambda_{3}/\lambda_{2})$--plane move towards the upper right corner.

The magnetic fluctuations of the dipole vortex are transverse, so the minimum variance direction $\mathbf{e}_3$ (when it is well-defined) 
is along the axis of the vortex. The angle between $\mathbf{e}_3$ and $\mathbf{B}_0$ is expected to be small $\theta_{B_0,3}\sim0^{\circ}$ according to the assumption of the model. Maximum and intermediate MVA eigenvectors $\mathbf{e}_1$, $\mathbf{e}_2$ lie in the plane perpendicular to $B_0$.

\subsection{Current sheets}

Current sheets are planar coherent structures that separate the plasma with different magnetic field directions. MHD classification of current sheets \correction{includes} rotational (RDs) and tangential (TDs) discontinuities \citep[e.g.,][]{baumjohann97,2011Tsurutani}. 

\subsubsection{Rotational discontinuity}

In Figure~\ref{fig:models} column (c) we show crossings of the RD model by a synthetic spacecraft (see details of the model in appendix~\ref{sec:RD}).
Rotational discontinuity has an arch-like hodograph (Figure \ref{fig:models}, column (c), third row).
Discontinuities with arch-shaped hodograph have been previously observed in the solar wind \citep{1989Neugebauer,Riley1996, 1996Tsurutani, 2010Sonnerup, 2012Haaland, 2013Paschmann}.
In the bottom panel, both ratios $\lambda_{2}/\lambda_{1} \simeq \lambda_{3}/\lambda_{2} \simeq 0$ when the noise level is low (see dots). For higher noise, 
$\lambda_{3}/\lambda_{2}$ increases more than $\lambda_{2}/\lambda_{1}$ (see crosses). 
If the noise level is small enough, so that the MVA eigenvectors $(\mathbf{e}_1,\mathbf{e}_2,\mathbf{e}_3)$ are well-defined, they coincide with the basis vectors $(\mathbf{y},\mathbf{z},\mathbf{x})$ of the reference frame of the sheet (for any crossing trajectory). 
\oa{The magnetic field magnitude is constant across the rotational discontinuity, so it is an incompressible structure.}

\subsubsection{Tangential discontinuity}

Figure~\ref{fig:models} column (d) shows the crossing of the tangential discontinuity. The details of the model are described in appendix~\ref{sec:TD}.
Independently on the crossing trajectory the polarisation is linear, $\lambda_{2}/\lambda_{1} \sim 0$, $ \lambda_{3}/\lambda_{2} \sim 1$. Only the maximum MVA eigenvector $\mathbf{e}_1$
is unambiguously defined, 
it is tangential to the discontinuity plane ($\mathbf{e}_1=\mathbf{y}$). 
The intermediate ($\mathbf{e}_2$) and minimum ($\mathbf{e}_3$) eigenvectors are in $x-z$ plane, where $\mathbf{x}$ and $\mathbf{z}$ are normal and guide-field directions correspondingly.
In this tangential discontinuity model $B_x=0$ $B_z=const$, so $\delta B_x=\delta B_z=0$. Consequently, MVA analysis can't distinguish between the normal ($\mathbf{x}$) and the guide field ($\mathbf{z}$) directions.
\oa{In general, the tangential discontinuity can be asymmetric, separating plasmas with different 
$|\mathbf{B}|$. Thus, the tangential discontinuity can be a compressible structure. \correction{Here, we use} 
the simple model, where the value of the magnetic field modulus $|\mathbf{B}|$ is the same on both sides of the discontinuity. }

\subsection{Magnetic holes}

Magnetic holes are compressible coherent structures, \correction{characterized} by a \correction{localized} decrease of the magnetic field modulus.  
Figure \ref{fig:models}, column (e) shows the crossing of the magnetic hole model. 
The details of the model and some observational properties of magnetic holes are discussed in the appendix~\ref{sec:Magnetic_holes}.
For any crossing trajectory, the polarisation is linear, $\lambda_{2}/\lambda_{1} \sim 0$, $\lambda_{3}/\lambda_{2} \sim 1$.

\section{Examples of the observed structures } \label{sec:structures}

We consider coherent structures detected by the integrated LIM over all scales and above the threshold, $I>I_{threshold}$ (see Section \ref{sec:LIM}). Among nearly $\sim 10^4$ events we have selected 374 with $I/I_{threshold}\ge 6$ for visual examination.  All of them have a \correction{localized} event at sub-ion scales, which is embedded in a larger event at ion scales. \correction{In} its turn, this ion scale event is embedded in an MHD scale event. This successive embedding, \correction{which} is shown here for the first time, is like the \correction{organization} of \textit{Russian dolls}.

\begin{figure*}[ht!]
	\centering
	\includegraphics[width=0.99\linewidth]{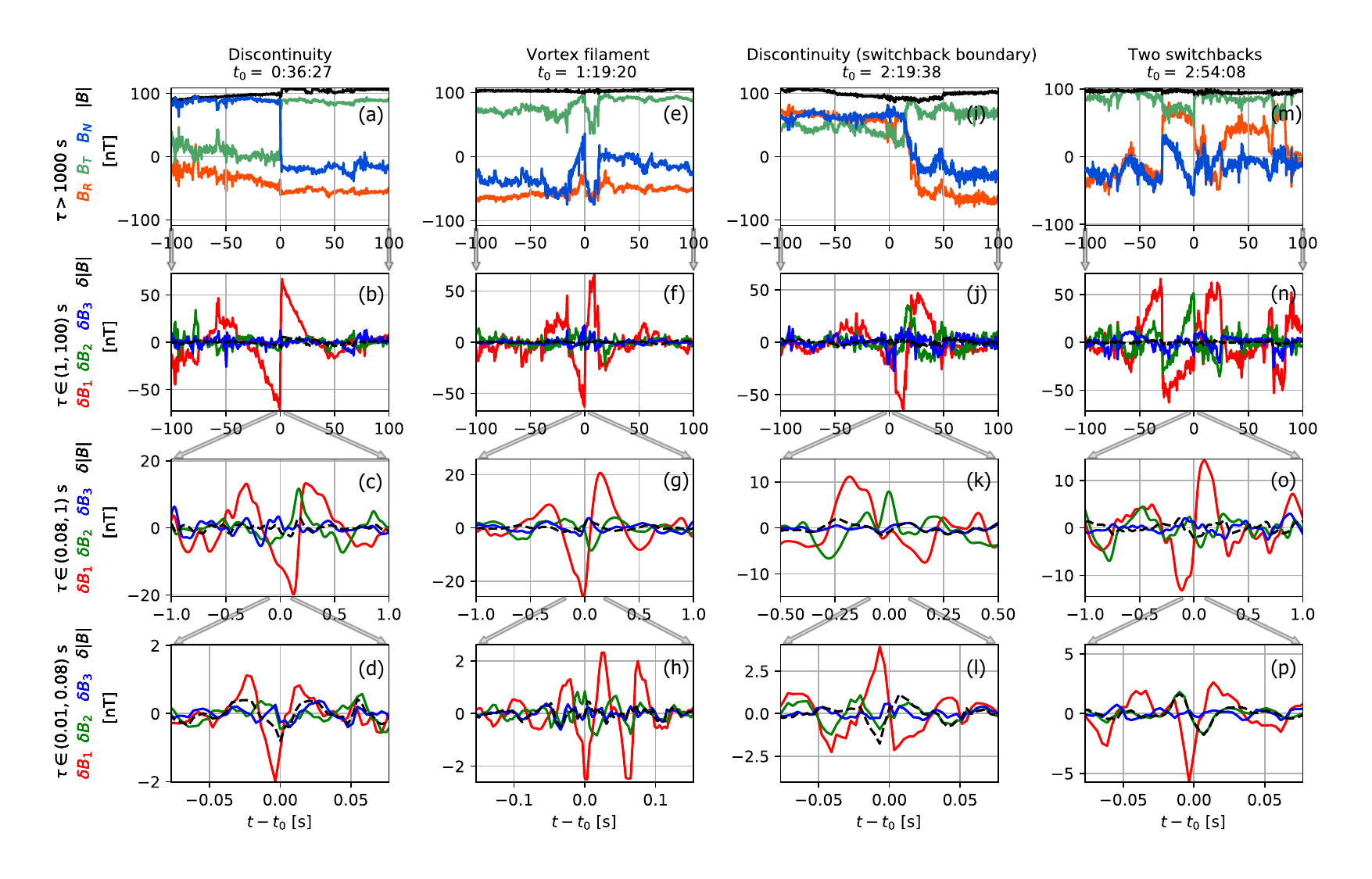}
\caption{Four examples of events detected on 6 November 2018 with integrated LIM $I/I_{threshold}>6$ are shown in columns. The central time of each structure $t_0$ is indicated in the title. 
Top row: Raw  magnetic field in \correction{the} RTN reference frame. 
Rows 2 through 4: the  bandpass filtered magnetic fluctuations at MHD, ion, and sub-ion frequency ranges in local MVA reference frames. 
At MHD scales these structures represent: a Tangential discontinuity (example 1), an Alfv\'en vortex (example 2), \oa{a rotational discontinuity at the switchback boundary} 
(example 3), and \oa{two neighboring switchbacks (example 4)}. In appendix~\ref{appendix:structures} the additional information (magnetic fluctuation polarisation, alfvenicity, plasma parameters) is provided for each event. This information \correction{complements} the interpretation of the coherent structures. At ion and sub-ion scales magnetic fluctuations can be interpreted as vortex-like structures for the 4 examples. } 
	\label{fig:FourExamples}
\end{figure*}
In Figure~\ref{fig:FourExamples} we show 4 such examples of different types of structures at different scales, found among the subset of 374 events. 
Here we restrict ourselves to show the magnetic fluctuations only.  
But the complementary information for each event is presented in appendix~\ref{appendix:structures}.
It shows the variation of the plasma parameters ($N_e,T_e,T_p,|V|$) across the structure and the alfv\'enicity.

We first consider the event in the left column of Figure~\ref{fig:FourExamples}~(a-d).
Panel~(a) shows the low-pass filtered magnetic field in \correction{the} RTN reference frame \correction{during $\pm100$~s}
around the central time with $t_0=[00:36:27]$~UT (on 6 November 2018).
\correction{Using the mean observed velocity \correction{$\sim 340$~km/s, time-scale of $200$~s corresponds to a space-scale $\sim 7\cdot 10^4$~km}.} 
In this case, the coherent structure, around $(t-t_0) = 0$ is associated \correction{with} a discontinuity but it is not associated \correction{with} a switchback boundary, since $B_R$ is not reversing.

In Figure~\ref{fig:FourExamples}~(b-d) we show the filtered magnetic field data $\delta B_j$ at time scales defined by Equation (\ref{eq:timescale_ranges}), with $j=$\textit{`MHD', `ion'} and \textit{`subion'}. 
We use \correction{the} local MVA reference frame  \citep[][]{MVA} adapted to each scale range shown. The basis vectors $(\mathbf{e}_1,\mathbf{e}_2,\mathbf{e}_3)$ are directed along the maximum, intermediate, and minimum variance of the magnetic field.

In Figure~\ref{fig:FourExamples}~(b), a high amplitude current sheet is observed at MHD scales. 
In appendix~\ref{sec:Example1} we use the plasma data to provide a physical analysis as summarized below.
In Figure~\ref{fig:structure1}~(i-k) we show that the Walen relation $\Delta \mathbf{V}=\pm \Delta \mathbf{B}/\sqrt{4 \pi \rho}$ for rotational discontinuities is violated in this example, and the pressure anisotropy correction \citep{Hudson1970} is insufficient to explain the observation.  
In addition, the magnetic field magnitude is changing across the sheet, see the black curve in Figure~\ref{fig:structure1}~(l). 
The \correction{normalized} amplitude of the jump is $\Delta B/B_{0}=0.1$, where $\Delta B=|B(t_{0}+\Delta t')|-|B(t_{0}-\Delta t')|=10$~nT, with $\Delta t'=3$~s, 
and $B_0=\langle|B|\rangle_{t-t_0\in(-100,100) \text{ s}}=100$ nT.
This \correction{localized} jump is $\approx 2$ times greater than the standard deviation of $\Delta B/B_{0}$ in the 5~h interval. 
This is also incompatible with the rotational discontinuity, across which the magnetic field magnitude is constant. 
\correction{In contrast, a change in the magnitude of the magnetic field is possible when crossing a tangential discontinuity.}
So we interpret this structure as a tangential discontinuity. 
The thickness of this current sheet can be estimated as  $d=V\Delta t=480$~km, or $100$~$\rho_i$ and $40$~$d_i$, where $\Delta t=1.5$~s is the duration of the current sheet crossing, local $V=340$~km/s.

Figure~\ref{fig:FourExamples}~(c,d) shows substructures at ion and sub-ion scales embedded in this tangential discontinuity. Ion scale structure, observed during $\Delta t=1$~s, resembles crossing the dipole Alfvén vortex model through its center, see Section~\ref{sec:Monopoles}. The cross-section scale is about $d=340$~km, or $60$~$\rho_i$ and $27$~$d_i$.
\oa{The structure, shown in the panel (d) is observed during $\Delta t=0.08$~s, so the cross-section scale is $d=27$~km$=2~d_i=5\rho_i$. This structure might represent a compressible ion scale Alfv\'en vortex \citep{Jovanovic2020}}.  
In this model, vortices have non-zero parallel magnetic fluctuations $\delta B_{\|}\ne 0$ maintained in pressure balance, and the $\delta B_{\perp}$ fluctuations are similar to the MHD Alfv\'en vortex.
The sketch illustrating the embedding is presented in Figure~\ref{fig:sketch}~(a). 

In Figure~\ref{fig:FourExamples}~(e-h) the second event  around $t_0=[01:19:20]$~UT is shown  in the same format as the first event. 
Radial magnetic field component $B_{R}$ is negative in the center of the structure and during the time interval $\pm 100$~s (MHD scales). 
So the MHD-scale coherent structure, shown in Figure~\ref{fig:FourExamples}(f), is not associated \correction{with} a switchback. Its time profile is consistent with the monopole Alfv\'en vortex crossing close to its center, see the black trajectory in Figure~\ref{fig:models}~(a).
The velocity and magnetic field fluctuations are correlated with a proportionality coefficient $\xi$, $\delta B/B_{0}=\xi  \delta V/ V_A$, where $\xi=0.86<1$, see appendix~\ref{sec:Example2}, Figure~\ref{fig:structure2}~(i-k). 
In the  model of Alfv\'en vortex, $\xi$ is a free parameter, simply related to other parameters of the vortex: $\xi=u/\alpha$, where $\alpha$ is the inclination angle \correction{of the vortex axis} with respect to the background magnetic field, and $u=V_{vortex}/V_{th,\perp}$ is the \correction{normalized} vortex propagation speed in the \correction{vortex plane.} 

The direction of the Alfv\'en vortex axis can be estimated using MVA if the vortex is crossed by the satellite off-\correction{center}. In this case, the vortex axis is directed along the direction of the minimum MVA direction $\mathbf{e}_3$.
In the considered example the Alfvén vortex is crossed near the \correction{center}, the direction of $\mathbf{e}_3$ (as well as $\mathbf{e}_2$) is not reliably determined. However, if we assume that $\mathbf{e}_3$ \correction{approximates} well the direction of the vortex axis, the inclination angle is $\alpha=4^{\circ}$. Then one can estimate the vortex propagation velocity $u=\xi\cdot \alpha=0.86\cdot(4/180\cdot\pi)=0.06$. 
\correction{Thus,} with \correction{a local proton thermal speed of} $v_{th,\perp}\simeq50$~km/s, the vortex propagation velocity \correction{is} $V_{vortex}=3$~km/s. This is negligible in comparison with the local bulk solar wind speed $V=330$~km/s, so the vortex is mainly convected across the satellite. 
Thus, we \correction{can} use the Taylor hypothesis to estimate its cross-section scale: $d=V\Delta t \sin\Theta_{BV}=2.4 \cdot 10^4$~km, or $5\cdot10^3 \rho_i$ and $2\cdot10^3 d_i$, where $\Delta t=80$~s is the duration of the structure \correction{and $\Theta_{BV}=70^{\circ}$ is the local field-to-flow angle}. 

Within this \correction{large-scale} monopole Alfv\'en vortex, 
we observe smaller embedded vortices, 
\correction{see Figure~\ref{fig:FourExamples}(g,h).}
The schematic sketch illustrating this embedding is presented in Figure~\ref{fig:sketch}(b). 
Note, that the sketch is not to scale, the diameters of the embedded vortices differ by a factor of 10: $d=440$~km $=90\rho_i =40d_i$ \correction{at scales that we call ion scales} and $d=47$~km $=9\rho_i = 4d_i$, \correction{at so-called sub-ion scales}. 

Figure~\ref{fig:FourExamples}(i-l) shows a third event around $t_0=[02:19:38]$~UT. 
In the panel (i), $B_R$ changes sign across the sheet, meaning that the sheet forms the boundary of a switchback \citep[see the complementary data in Figure~\ref{fig:structure3}][and Appendix~\ref{sec:Example3}], similarly to observations by \citet{2020Krasnoselskikh}.
The current sheet is located at $t-t_{0}=20$~s, \correction{see Figure~\ref{fig:FourExamples}(j)}.
The velocity and magnetic field jumps satisfy the Walen relation and the magnetic field magnitude is constant within the short interval $t-t_{0} \in (0,40)$~s near the center of the sheet. So we conclude that this discontinuity is rotational. \correction{The thickness of the current sheet ($\sim 40$~s), in terms of spatial scales, corresponds to $1.4\cdot 10^4$~km $=2\cdot 10^3 \rho_i = 1.2 \cdot 10^3 d_i$.} 

Ion scale magnetic fluctuations are shown in Figure~\ref{fig:FourExamples}(k).  
The minimum variance direction $\mathbf{e_3}$ is along the local mean magnetic field $\mathbf{B}_0=\langle\mathbf{B}\rangle_{t-t_0\in(-0.5,0.5) \text{ s}}$. 
The maximum ($\delta B_1$) and intermediate ($\delta B_2$) MVA components are perpendicular to $B_0$ and have similar amplitudes. At the center of the structure ($t-t_0=0$) $\delta B_1$ changes sign, and $\delta B_2$ has a peak. 
So, magnetic field fluctuations are in accordance with the off-center crossing of the monopole Alfv\'en vortex, see \correction{the} blue trajectory in Figure~6(a). 
The cross-section scale of the vortex is $150$~km, or $\sim 25\rho_i$ and $12d_i$.

Sub-ion scale structure, Figure~\ref{fig:FourExamples}~(l), has typical properties for structures at sub-ion scales in our statistics: $\delta B_{1}$ has a Mexican hat-like shape, and this event has a significant compressibility $\delta |B| \sim0.5 \, \delta B_{1}$. It is very similar to the sub-ion scale structure shown in Figure~\ref{fig:FourExamples}~(d). 
The cross-section scale of this structure is 24~km, or 4~$\rho_i$ and 2~$d_i$. Such localized compressible magnetic fluctuations at ion scales can be interpreted as the ion Alfv\'en vortex of \citet{Jovanovic2020}.  

Figure~\ref{fig:FourExamples}~(m-p) shows the fourth example around $t_0=[02:54:08]$~UT. Our detection method \correction{catches} two neighboring switchbacks, lasting 200 seconds at MHD scales. Indeed, we see 
in panel (m) that $B_R>0$ during two time intervals in the center, and the rest of the time $B_R<0$. Figure~\ref{fig:FourExamples}(n)
\correction{shows} four current sheets at the boundaries of these switchbacks.  
The central times of the four current sheets are $t-t_0=\{-29,0,25,85\}$~s. 
The complementary information is presented in appendix~\ref{sec:Example4}. 

At ion scales, Figure~\ref{fig:FourExamples}(o), we observe an embedded coherent structure that might represent a monopole Alfvén vortex crossed through the center. The cross-section scale is $d=160$~km $=23\rho_i=13d_i$. We observe a strong peak-like fluctuation of $\delta B_1$ at sub-ion scales (see panel (p)). The intermediate MVA fluctuation $\delta B_2$ is \correction{localized} in the center of the event. The profile of $\delta B_2$  is closely similar to the one of $\delta |B|$. So this structure is compressible $\delta |B| \sim 0.2 \, \delta B_{1}$. The cross-section scale is $d=12$~km $\simeq 1.7\rho_i = d_i$. It belongs to the same typical class of sub-ion scale coherent structures (compressible ion scales Alfv\'en vortices) as in other examples.

\section{Multiscale minimum variance analysis} \label{sec:minvar}

\begin{figure}[ht!]
	\centering
	\includegraphics[width=.8\linewidth] {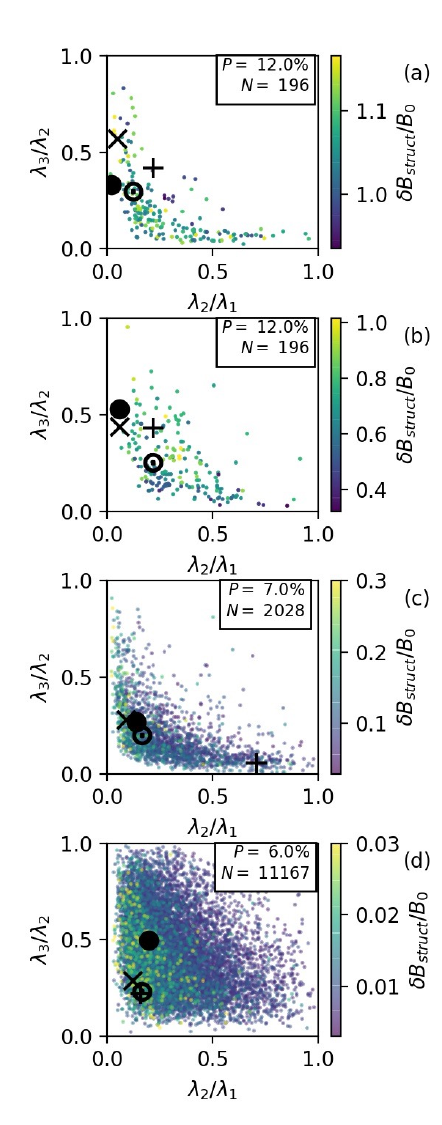}
\caption{Minimum variance analysis eigenvalues ratios plane ($\lambda_{2}/\lambda_{1},\lambda_{3}/\lambda_{2}$): each dot corresponds to an observed coherent  structure, the color gives its amplitude $\delta B_{struct}/B_0$ (see Equation~\ref{eq:deltaB}). 
Panels (a,b) correspond to the raw data and MHD scales, respectively. They include 196 structures found at MHD scales. Panel (c) gives the results for 2028 structures at ion scales and panel (d) gives the eigenvalues ratios for 11167 events at sub-ion scales. 
The filling factor $P$ and the number of detected coherent structures $N$ at different frequency ranges are shown in the legends. The eigenvalue ratios of the example structures 1-4 from the Figure~\ref{fig:FourExamples}
are shown by the black marks: "circle", "cross", "plus", and "odot".} 
	\label{fig:ratios}
\end{figure}

Now, we consider the whole set of structures detected by integrated LIM at different time-scale ranges, see Equation~(\ref{eq:I_j}). As we have discussed in Section~\ref{sec:LIM}, the number of structures increases toward small scales, from nearly 200 events at MHD scales to more than $10^4$ events at sub-ion scales.  For all these events, we study the amplitude anisotropy of the measured fluctuations via \oa{Minimum Variance Analysis \citep[MVA,][]{MVA}}. Then, we compare the observed anisotropy with one of the model structures crossed by a spacecraft.

\subsection{Observational characteristics of coherent structures}

For each coherent structure detected at j-th range of scales we consider filtered magnetic field fluctuations $\delta \mathbf{B}_{j}$ at the time interval $t-t_{0} \in T_{struct}= (-\tau_{max,j},\tau_{max,j})$ in the vicinity of the structure center $t_{0}$, where $\tau_{max,j}$ is the maximum timescale of each scale range defined by Equation~(\ref{eq:timescale_ranges}).
We define the amplitude of the structure $\delta B_{struct,j}$ as:
\begin{equation}   \label{eq:deltaB}
\delta B_{struct,j} = \max(|\delta \mathbf{B}_{j}|)_{t \in T_{struct}}
\end{equation}

The amplitude anisotropy of the magnetic fluctuations $\delta \mathbf{B}_{j}$ of the structure along the crossing trajectory is \correction{characterized} by MVA eigenvalue ratios  $\lambda_{2}/\lambda_{1}$ and $\lambda_{3}/\lambda_{2}$.
The relative amplitude $\delta B_{struct,j}/B_0$ is shown in color in the Figure~\ref{fig:ratios}. For each range of scales, the number of structures $N$ and the filling factor $P$ \correction{(defined in the same way as eq.~\eqref{eq:filling_factor} but for each scale range)}  are shown in the legend.

Figure~\ref{fig:ratios}(a) gives the results of the MVA for the raw magnetic field data during 200~s time intervals around the central times $t_0$ of the MHD-scale coherent structures (see the discussion of the detection method \correction{at} the end of Section~\ref{sec:LIM}). 
The MVA results for four 
examples \correction{analyzed} in detail in Section~\ref{sec:structures}
are marked on the $(\lambda_{2}/\lambda_{1}$,$\lambda_{3}/\lambda_{2})$ plane with special symbols:  Example-1, TD at large scales, is a black dot; Example-2, an Alfv\'en vortex at large scales, is a cross; Example-3, a RD at large scales, is a plus; Example-4, two neighbouring switchbacks, is a circled dot.

For a large number of events\correction{,} the ratio of the minimum over intermediate eigenvalues is small $\lambda_3/\lambda_2<0.5$, while intermediate over maximum variance, $\lambda_2/\lambda_1$, takes the whole range of values from 0 to 1.
Among the considered models, this zone on the eigenvalue plane corresponds only to the monopole and dipole Alfv\'en vortices, see Figure~\ref{fig:models}.
Minimum over intermediate variance, $\lambda_3/\lambda_2$, \correction{sometimes} takes high values ($>0.5$), as is the case for the monopole vortex, a tangential discontinuity\correction{,} or a magnetic hole.  Values of $\lambda_3/\lambda_2$ around 0.3 and for small $\lambda_2/\lambda_1$ can be interpreted as rotational discontinuities, see Figure~\ref{fig:models}. 
So, the observed distribution of $\lambda_3/\lambda_2$ as a function of $\lambda_2/\lambda_1$ can be due to a superposition of different types of coherent structures. It seems that vortices are dominant, but other types of structures may also exist. 

Figure~\ref{fig:ratios}(b) corresponds to the same set of coherent structures as in panel (a) but for filtered MHD-scale fluctuations $\delta \mathbf{B}_{MHD}$  
instead of the raw magnetic field data. 
Here, the data are spread nearly uniformly in the bottom-left part of the panel. This distribution can be also interpreted as a superposition of the 5 models discussed above, with a dominance of vortices.

Figure~\ref{fig:ratios}(c,d) represent the MVA results for ion and sub-ion scale structures, respectively. 
At ion scales, the distribution is similar to what is observed in raw data, but with more cases (2028 vs 196).
Sub-ion scale structures have different \correction{distributions} on the MVA eigenvalue ratios.  
Most of the points \correction{(especially yellow ones, corresponding to high-amplitude events)} are grouped closer to the left part of the eigenvalue plane, where $\lambda_{2}/\lambda_{1}<0.25$. But this does not exclude any of the 5 models.

Below, in Sections~\ref{sec:Noise_estimation} and \ref{sec:classification}, we propose a new systematic approach to quantify the proportions of different types of structures at different scales. 

\subsection{Noise level estimation}\label{sec:Noise_estimation}

We want to compare the observed distributions of $(\lambda_{2}/\lambda_{1},\lambda_{3}/\lambda_{2})$ and the degree of compressibility (defined below in equation (15)) 
for MHD, ion and sub-ion scale structures,
with the crossings of different coherent structures models, (see Section~\ref{sec:models}). 
The incoherent noise affects the MVA eigenvalue ratios (shown in the bottom row of Figure~\ref{fig:models}). The greater is the ratio $\epsilon=\delta B_{noise}/ \delta B_{struct}$, the closer are the 
$\lambda_{2}/\lambda_{1}$ and $\lambda_{3}/\lambda_{2}$ to 1.
Therefore, we need to estimate $\epsilon$ from observations to take into account the noise in the model crossings.

For each structure at j-th scale range, we calculate the ratio of the noise $\sigma_{noise,j}$, defined in Equation (\ref{eq:sigma_noise}), to the amplitude of the structure $\delta B_{struct,j}$: 
  \BE \label{eq:epsilon_struct}
  \epsilon_{obs,j} = \sigma_{noise,j}/\delta B_{struct,j}
  \EE
At each range of scales the distribution of $\epsilon_{obs,j}$ is nearly Gaussian, but with different values of parameters. The mean values $\langle \epsilon_{obs,j} \rangle$ and the standard deviations $\sigma( \epsilon_{obs,j})$ are shown in the Table \ref{table:2}. 

\begin{table}[t!]
\centering
\begin{tabular}{lll}  
  &$\langle \epsilon_{obs} \rangle$& $\sigma( \epsilon_{obs,j})$\\
\hline
\hline
RAWDATA MHD &0.11&0.03 \\
MHD         &0.11&0.03 \\
Ion scales  &0.15&0.05 \\
Sub-ion     &0.12&0.03 \\
\end{tabular}
\caption{The mean and the standard deviation of the relative noise level $\epsilon_{obs}$ at different ranges of scales (defined in Equation (\ref{eq:epsilon_struct})).
}
\label{table:2}
\end{table}

We repeated the crossings simulation with 10 different relative amplitudes of the imposed noise $\epsilon_{sim}$ following the Gaussian distribution with the same parameters, $\langle \epsilon_{obs,j} \rangle$ and $\sigma( \epsilon_{obs,j})$, as in observations. The obtained results of the model crossings with different $\epsilon_{sim}$ are used in the next Section.

\begin{table*}
\centering
\begin{tabular}{lllllllll}  
  &&& \multicolumn{2}{c}{Alfvén vortex} & \multicolumn{2}{c}{Current sheet} & \multicolumn{1}{c}{Magnetic hole}& \multicolumn{1}{c}{None}\\
  &$N$& $P$ [\%] & \multicolumn{1}{c}{Monopole} & \multicolumn{1}{c}{Dipole}
  & \multicolumn{1}{c}{Rotational}& \multicolumn{1}{c}{Tangential} & & \\
\hline
\hline
RAWDATA MHD &196 &12& 0.04 &  0.86  &  0.1  & 0    & 0 & 0  \\
MHD         &196 &12& 0.1  &  0.84  &  0.0  & 0    & 0 & 0.06 \\
Ion scales  &2028 &7& 0.15 &  0.85  & 0.0   & 0  & 0  & 0 \\
Sub-ion     &11167&6& 0.07 &  0.49  & 0.05  & 0  & 0.004  & 0.39  \\
\end{tabular}
\caption{In the 1st and 2nd columns, we give the number of structures $N$ and the filling factor $P$ [\%] at different ranges of scales (as defined in Section~(\ref{sec:LIM})). Other columns give results of the problem formulated in Equation~(\ref{eq:Pmodel}): the coefficients $p_{mod}$  correspond to the fraction of the observed coherent structures that have MVA eigenvalue ratios and compressibility consistent with the crossing of a given model (Figure~\ref{fig:models}). 
}
\label{table:3}
\end{table*}

\begin{table*}
\centering
\begin{tabular}{lllllll}  
    & \multicolumn{2}{c}{Alfvén vortex} & & \multicolumn{2}{c}{Current sheet} & \multicolumn{1}{c}{Non-identified} \\
    & \multicolumn{1}{c}{Isolated} & \multicolumn{1}{c}{Multiple} & \multicolumn{1}{c}{Vortex+Current sheet}& \multicolumn{1}{c}{Isolated}& \multicolumn{1}{c}{Non-isolated} & \\
\hline
\hline
RAWDATA \& MHD & \oa{0.24 (0.03)} &  \oa{0.16 (0.03)}  &  \oa{0.26 (0.08)}  & \oa{0.03 (0.005)} & \oa{0.10 (0.05)}  & \oa{0.21} \\
\end{tabular}
\caption{Fraction of the MHD structures obtained by visual classification of 196 events at these scales.  We give the fraction of the switchbacks  in the  parenthesis. In  total, 19\% of events at MHD scales are parts of  switchbacks.
}
\label{table:4}
\end{table*}

\subsection{Classification}\label{sec:classification}
For convenience we use the notation $(r_{21},r_{32})=(\lambda_2/\lambda_1,\lambda_{3}/\lambda_{2})$ for MVA eigenvalue ratios. 
First, we investigate the presence of magnetic holes.
We use two criteria to select magnetic holes: high compressibility and linear \correction{polarization}.

A coherent structure is compressible if the magnetic field modulus $|\mathbf{B}|$ is not constant because of the parallel magnetic fluctuations of the structure.
Considering the compressibility at j-th range of scales, we filter $|\mathbf{B}|$ (as we do for fluctuations $\delta \mathbf{B}_j$) to define $\delta |\mathbf{B}|$ at the scale-range $j$. 
The amplitude of compression associated with a coherent structure is given as $\max(|\delta |\mathbf{B}||)_{t \in T_{struct}}$. We normalise it by $\delta B_{struct}$ to define the compressibility of the structure:
   \begin{equation}   \label{eq:compressibility}
   C_{struct}=\max(|\delta|B||) /\delta B_{struct}
   \end{equation}
We underline that our definition of compressibility differs from the definitions used in \cite{1977Turner,2020Volwerk}. It is more similar to those used in \cite{2007Stevens,2016Perrone}.

First, we impose $C_{struct}>0.8$ to select strongly compressible structures and second, we delimit the zone $(r_{32}>0.6,r_{21}<0.4)$ in the MVA eigenvalue ratios plane, that is characteristic for the magnetic hole crossings, see the bottom panel of the Figure~\ref{fig:models}~(e). \oa{This zone is a bit wider than in Figure~\ref{fig:models}~(e), because for some of \correction{the} magnetic holes in the observational statistics, the relative noise amplitude could be higher than $\epsilon=0.1$, as used in the model.}
The percentage of MHD, ion, and subion structures satisfying both criteria is presented in the  column \textit{Magnetic hole} of Table~\ref{table:3}.
\oa{We found that magnetic holes are detected only at sub-ion scales.} Among sub-ion scale structures, they account for 0.4\% of the cases. We will study these events in more \correction{detail} in future work.

We define the proportions of vortices and current sheets among the remaining observed structures by comparing the amplitude anisotropy from observation, without imposing any criterion for compressibility.

\begin{figure*}[ht!]
	\centering
	\includegraphics[width=1.01\linewidth]{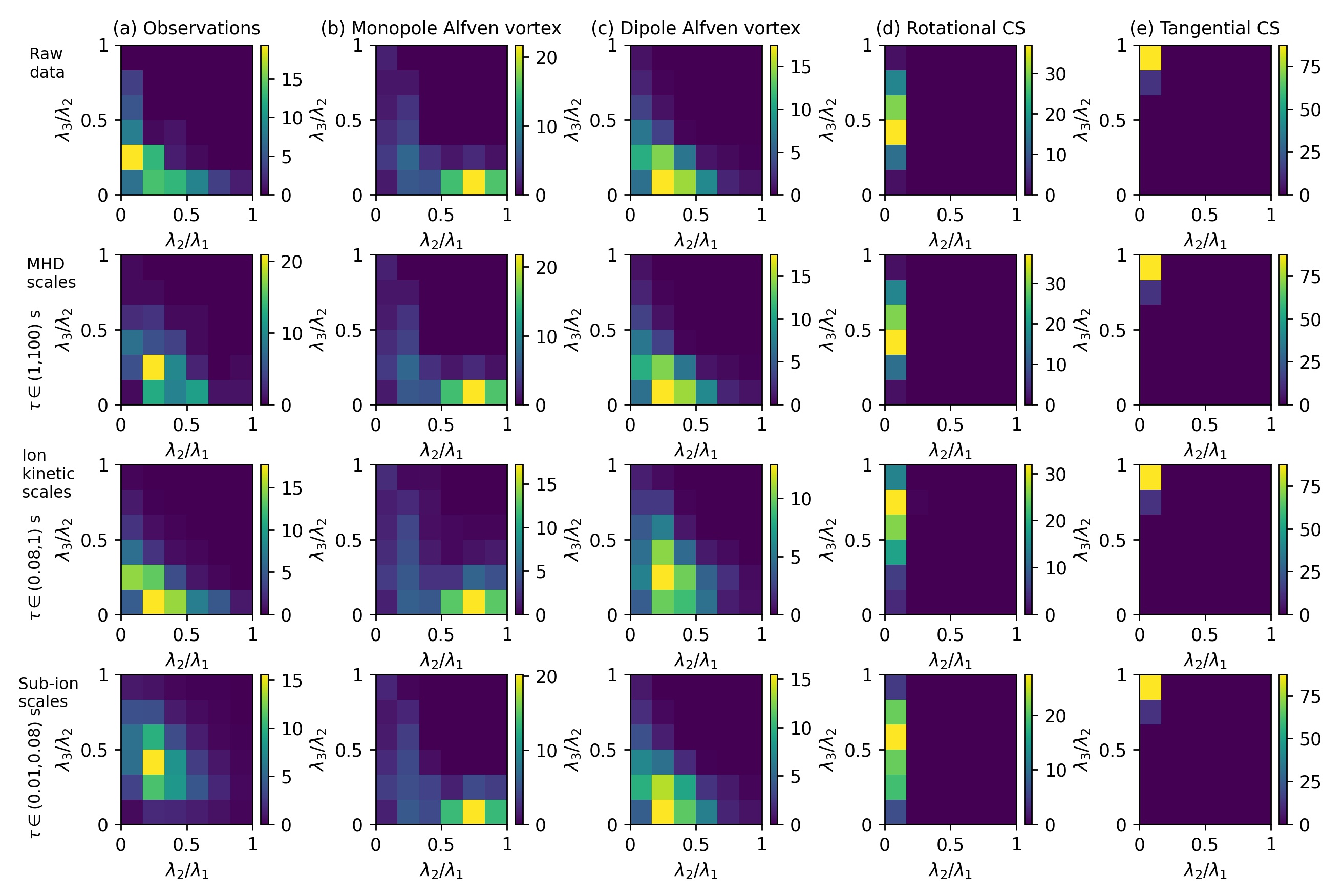}
\caption{Probability distributions on the MVA eigenvalue ratios plane $(r_{21},r_{32})=(\lambda_2/\lambda_1,\lambda_{3}/\lambda_{2})$. The column (a) shows the probability~[\%] per bin to observe a coherent structure with the corresponding MVA eigenvalue ratios  (so $P_{obs,j}(r_{21},r_{32})$ defined by Equation (\ref{eq:Pij_obs})). \correction{The} first and second panels of column (a) show the distributions for the MHD-scale coherent structures using the raw (non-filtered) data and the MHD-range filtered data respectively. \correction{The} third and fourth rows of column (a) correspond to coherent structures detected at ion and sub-ion scale ranges.
Columns (b-e) show the probability densities obtained from simulating model crossings ($P_j(r_{32},r_{21}|\text{model})$ defined by Equation (\ref{eq:Pij_model})). The difference between panels of different rows (b-e) is due to the different imposed noise \correction{levels} $\epsilon_{sim}$ (see Section \ref{sec:Noise_estimation} for details). 
} 
	\label{fig:Pij}
\end{figure*}

Figure~\ref{fig:Pij}(a) \correction{shows} 2D histograms ($6 \times 6$ bins) of distributions of the data in $(r_{21},r_{32})$--plane for observations at MHD (top), ion (middle) and sub-ion (bottom) 
scales. In other words, we show the probability density $P_{obs,j}$ of observations 
  \BE \label{eq:Pij_obs}
  P_{obs,j}(r_{21},r_{32})=N_{obs,j}(r_{21},r_{32})/N_{obs,j},
  \EE
where $N_{obs,j}(r_{21},r_{32})$ is the number of the observed structures in a bin, and $N_{obs,j}$ is the total number of observed structures. The index $j$ denotes the scale range. 

We assume that crossings of coherent structures along trajectories with different impact parameters are equally probable and we take into account the noise from the observations, with Equation~(\ref{eq:epsilon_struct}), 
as explained below.  
Since the dipole Alfv\'en vortex has an angular structure, we average the results over a uniform distribution of trajectory orientations. Then, we obtain the probability density $P(r_{21},r_{32}|\text{mod})$ of MVA eigenvalue ratios for each model structure: 
  \BE \label{eq:Pij_model}
  P_j(r_{21},r_{32}|\text{mod})=N_{mod,j}(r_{21},r_{32})/N_{mod,j}
  \EE
The probability distributions for 4 different models  $P_j(r_{21},r_{32}|\text{mod})$ are shown in columns (b-e) of Figure~\ref{fig:Pij}. 
To simulate different scales, we change the \correction{noise level} according to the estimated value at each scale, see Equation~(\ref{eq:epsilon_struct}).

The observed distribution of MVA eigenvalue ratios $P_{obs,j}$ can be expressed as the linear combination of the conditional probabilities $P_j(r_{21},r_{32}|\text{mod})$, determined from the models. The positive coefficients $p_{\text{mod}}$
reflect the probability to encounter each model structure. Coefficients $p_{\text{mod}}$ are found from the constrained minimization problem:
\begin{equation}\label{eq:Pmodel}
    \begin{cases}
    ||P_{\text{obs}}(r_{21},r_{32})-\displaystyle\sum_{\text{mod}}p_{\text{mod}} \, P(r_{21},r_{32}|\text{mod})||\to 0\\
    \displaystyle\sum_{\text{mod}} p_{\text{mod}}\le 1\\
    p_{\text{mod}}\ge 0
    \end{cases}
\end{equation}
\correction{where $|| . ||$ means a norm (the square root of the square of the difference of the matrix)
that allow us to use the least squares minimization.}
The problem is solved for each range of scales. For convenience\correction{,} the index $j$ is omitted in Equation~\eqref{eq:Pmodel}.

The resulting probabilities $p_{\text{mod}}$ 
are shown in the Table~\ref{table:3}.
For each range of scales\correction{,} the observations 
are most consistent with the crossings of the dipole Alfv\'en vortices ($\ge 80\%$). The monopole vortices account for \correction{a small fraction of coherent structures, $7-15\%$ depending on the scale range}. 
The rotational discontinuities are observed in raw (non-filtered) data at MHD scales only, 
but not in the MHD band-pass filtered data.  
\correction{This is because} the band-pass filter makes the waveforms very different from the current sheet simple model, see Figure~7(b).
Tangential discontinuities \correction{do} not appear to be statistically significant. 
There is $6\%$ of events which were not possible to model at MHD scales and $39\%$ at sub-ion scales (see the \textit{None} column in  Table~\ref{table:3}). These unidentified large number of events at sub-ion scales is probably due to a more 3-\correction{dimensional} nature of the fluctuations not taken into account by nearly incompressible models \oa{with mostly $\delta B_{\perp}$ fluctuations}. 

The result presented in Table~\ref{table:3} does not change qualitatively if\correction{,} instead of least squares, the sum of the absolute values of probability differences (between observations and models) in each bin is minimized.

\correction{Let us compare} the obtained results 
with the previous observations. 
The visual classification of ion-scale coherent structures at 0.17 au, during the first PSP perihelion, has been done recently in \cite{2020Perrone}. 
Three different time intervals were considered: quiet, weekly-disturbed\correction{,} and highly-disturbed solar wind. The highly-disturbed interval (of 1.5 h) with $B_R$ reversals is a subset of the 5h-interval considered here. 
The authors concluded that in the highly-disturbed interval current sheets were dominant ($46$\%), while during the weekly-disturbed interval Alfvén vortices ($45$\%) and wave packets ($50$\%) were observed.
This is in contrast with the quantitative classification results \correction{obtained here} at ion scales, showing that Alfvén vortices are dominant.

In the previous studies of ion scales coherent structures at 1~au in slow \citep{2016Perrone} and fast \citep{2017Perrone} solar wind with Cluster satellites, the dominance of Alfv\'en vortices with respect to current sheets has been found. \correction{This} is more consistent with our results at 0.17~au in the slow wind.

We have visually \correction{analyzed} 196 MHD-scale coherent structures to understand why the percentage of current sheets is low. \correction{We remind that the peaks of the integrated LIM over the MHD range (between 1 and 100~s) determine the central times of the structures.} The time interval for each structure is $\pm100$~s around the central time, the corresponding spatial scale is $\sim 7\cdot10^{4}$~km. 
The results are summarized in the Table~\ref{table:4}: We \correction{realized} that the observed events need to be separated in isolated and non-isolated structures, like a train of Alfv\'en vortices (see the second column `Multiple Alfven vortices') or non-isolated current sheets (column 5) and as well we observe a \correction{co-existence} of a large scale weak current sheet with a vortex inside (column 3). 
So, in comparison with an automatic classification (Table~\ref{table:3}), here we observe more current sheets, most of them are associated with a vortex (51 cases out of 196), but as well 6 isolated current sheets which have been detected in Raw data but not at MHD scales.  For the non-isolated current sheets and especially those \correction{associated} with a vortex, the automatic classification tends to interpret the structures as Alfv\'en vortices.

Indeed, if the current sheet is not isolated,
the perturbations of the neighboring structure affect the MVA eigenvalue ratios, so that the structure is shifted away from the characteristic zone on the $(r_{21},r_{32})$--plane, see Figure~\ref{fig:models}.
Consequently, these events contribute to $p_{monopole}$ and $p_{dipole}$ in the solution of the minimisation problem (Eq.~\ref{eq:Pmodel}).

Finally, for the 196 events, we have visually \correction{inspected} ion and sub-ion scale sub-structures. Their shape is consistent with vortex crossings and not with current sheets. \correction{This} is in agreement with  Table~\ref{table:3}.

\section{Conclusion and discussions} \label{sec:conclusion}
The intermittency in the solar wind is typically investigated from the statistical point of view. The scale-dependent kurtosis of magnetic increments is used as the principal quantitative diagnostic, showing the presence of coherent structures. 

\begin{figure}[ht!]
	\centering
	\includegraphics[width=1\linewidth]{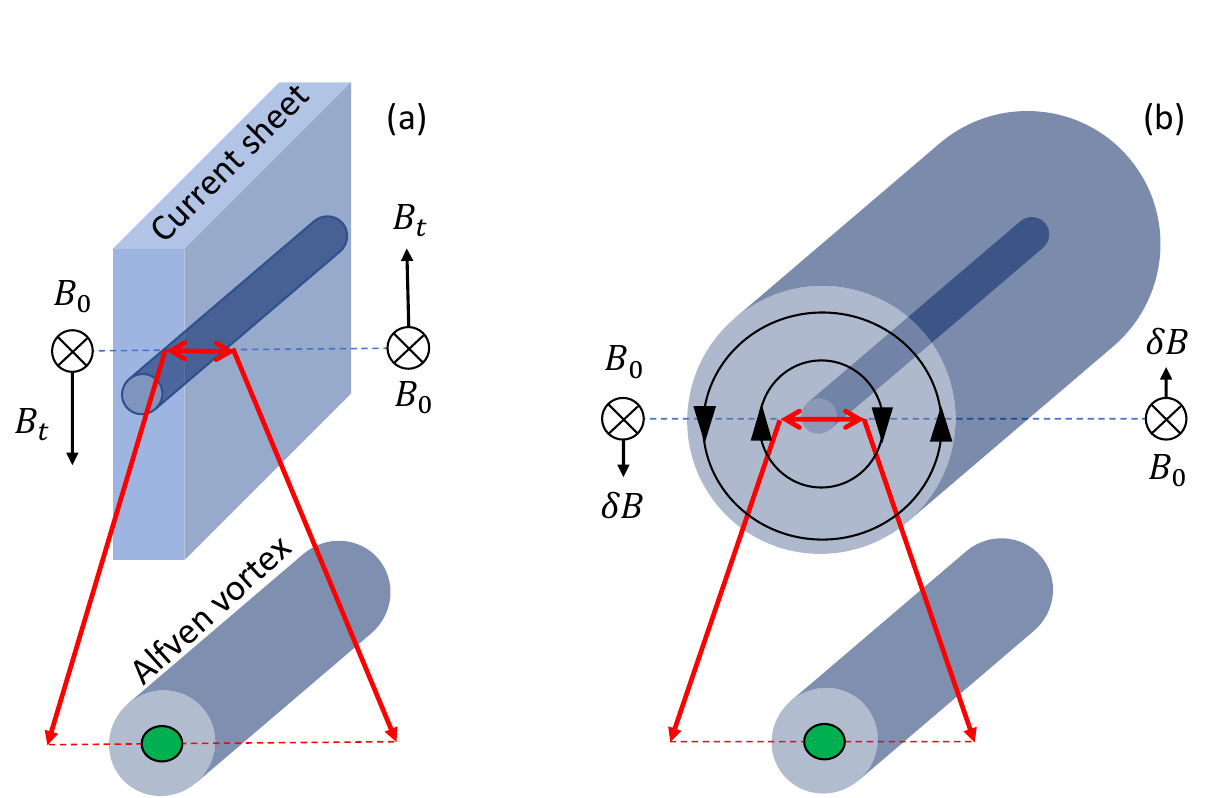}
\caption{Panel~(a): Schematic sketch of the Example 1, shown in the Figure \ref{fig:FourExamples}~(a-d). The blue dashed line illustrates the crossing trajectory. The mean magnetic field ($B_0$) and the tangential ($B_t$) component are shown in black on both sides of the current sheet. The red lines indicate the zoom to the embedded ion-scale vortex. A compressible sub-ion scale vortex is shown in green. The embedded sub-structures are shown not to scale.
Panel~(b): sketch of the Example~2, shown in the Figure~\ref{fig:FourExamples}~(e-h). The magnetic field fluctuations associated with a vortex are shown in black. The embedding of ion and sub-ion scale vortices is shown in the same format as in panel~(a).} 
	\label{fig:sketch}
\end{figure}

In this paper, for the first time, we apply a multi-scale approach in physical space, from the largest MHD scales $\sim 10^5$~km to the smallest resolved sub-ion scales  $\sim 3$~km.

Using PSP merged magnetic field data at 0.17~au and the Morlet wavelet transform, we detect intermittent coherent structures covering all scales.
We find \correction{localized} magnetic fluctuations on subion scales with amplitudes up to $\delta B/B_0\sim 0.03$, where $B_0$ is the local magnetic field strength. These small-scale structures are typically embedded in a larger structure at ion scales, with amplitudes up to $\delta B/B_0\sim 0.3$. At its turn, \correction{the} ion scale structure is embedded in a high-amplitude ($\delta B_{struct}/B_0\sim 0.4-1.0$) MHD-scale structure.
Such embedding across the whole turbulent cascade is presented here for the first time.

The topology and properties of the coherent structures change from scale to scale. 
Using plasma and magnetic field time profiles, we characterize several events in more \correction{detail}.
We show examples of planar \correction{tangential and rotational} discontinuities at MHD scales with the thickness of about $100\rho_i$ containing  embedded sub-structures inside it: incompressible ion-scale Alfv\'en vortex \oa{(with a cross-section of $\simeq 30\rho_i$)} and sub-ion scale compressible vortex \oa{$\simeq 5\rho_i$}, see the sketch in Figure~10(a). 
Another example is an Alfv\'en vortex at MHD scales \oa{with the cross-section of $5\cdot 10^3\rho_i$}, with embedded incompressible vortices \oa{of $60\rho_i$} at ion and \oa{of $\sim 10\rho_i$ at} sub-ion scales, see the sketch in Figure~10(b).

We completed the study of examples with a statistical analysis.
In a time interval of about 5 hours we detected nearly 200 events at the MHD scales, and \correction{many} more events at ion ($\sim 2\cdot 10^3$) and sub-ion scales  ($\sim 10^4$). 
The filling factor of the structures, that we estimate in a conservative way\footnote{Using only the time where the integrated LIM is over the threshold means that the lower-energy part of each event is not taken in-to account in the calculation of the filling factor. If the method of \citep{2016Perrone} was used, the filling factor would be 2-3 times larger. } (see discussion in section~4.1), decreases from 12\% at MHD scales to 7\% and 6\% at ion and sub-ion scales correspondingly. 

To determine the dominant type of coherent structures, we perform an automatic  classification based on the comparison of the observed amplitude anisotropy
of magnetic fluctuations within the observed events at all scales with analytical models of Alfvén vortices, current sheets\correction{,} and magnetic holes.
We do not consider magnetosonic shocks as far as the \correction{analyzed} time interval is mostly incompressible at MHD scales.
The results show the dominance of Alfv\'en vortices at all scales and only \correction{a} few current sheets, \correction{and} mostly in raw data.

In order to understand the low number of current sheets, we did a visual inspection of 196 events at MHD scales. It reveals that isolated current sheets are indeed rare (3\%). 
Most of the events are rather complex. 
About 10\% of structures represent non-isolated current sheets. Many of the detected events are vortices within the current sheets (26\%). Isolated and non-isolated vortices account for 66\% of the structures. In all these cases, except isolated current sheets, the automatic classification method \correction{tends} to interpret the data as vortices. 

\correction{The} ion transition range of scales corresponds to $[30,340]$~km, or $[5,70]\rho_i$ in our time interval. Here, with the automatic classification method\correction{,} we found 85\% dipole vortices and 15\% of monopoles. 
Planar discontinuities are not found by our method.
Visual inspection of 196 cases confirms the absence of current sheets at these scales.

On sub-ion scales ($[3,30]$~km or $[0.5,5]~\rho_i$) coherent structures represent dipole vortices (49\%), monopole vortices (7\%), rotational current sheets (5\%) and magnetic holes (0.4\%). Around 39\% of sub-ion scale structures do not fit any of the considered models. 
It is, \correction{possibly,}
because the incompressible models of vortices have been used \correction{to} compare \correction{with the} observations. To improve this study at these scales in the future, the compressible ion scales Alfv\'en vortex model of \citep{Jovanovic2020} should be used. 
In this model, together with typical vortical $\delta \mathbf{B}_{\perp}$, the compressible component $\delta B_{\|}$ appears, which is in pressure balance with density fluctuations. We presume, that the comparison with this model will increase the proportion of the vortices at sub-ion scales.

Results presented in this article show the dominance of Alfv\'en vortices at all scales
\correction{: from inertial to sub-ion range.}
Thus, Alfv\'en vortices are important building blocks of solar wind turbulence \correction{and in particular of its intermittency at all scales.} Alfv\'en vortices \correction{can} explain $\delta\mathbf{V}_{\perp}/V_A=\xi \,\delta\mathbf{B}_{\perp}/B_0$, with $\xi \ne 1$, typically observed in the solar wind for Alfv\'enic periods.

\correction{In the reflection-driven turbulence in  the reduced magnetohydrodynamic  numerical simulation of \citet{Meyrand2023}, the authors observe \textit{cellularisation} of turbulence with generation of magnetic vortices with $\delta \mathbf{V}_{\perp}=0$, with $k_{\|} = 0$ and with magnetic field discontinuity at vortex boundary.  
These structures are thus quite different from the smooth Alfv\'en vortices observed here with $\delta \mathbf{B}_{\perp}\sim \delta \mathbf{V}_{\perp}$ correlations, but they have similar twisted magnetic field configuration and $k_{\|} = 0$.}

Our results are limited to a specific slow highly-perturbed solar wind region at 0.17~au from the Sun. The analysis can be expanded to different solar wind conditions (different radial \correction{distances}, types of solar wind, originating from ecliptic or polar regions of the Sun) to obtain a more general picture.

\correction{The} multiscale nature of coherent \correction{structures} described in this article can be studied in \correction{the} future by the Helioswarm (NASA mission). It will cover MHD, ion, and sub-ion scales at the same time.

\acknowledgements
\oa{We thank the reviewer for his/her careful reading and analysis of our paper. The comments were very helpful to improve the paper.}
AV acknowledges funding support from the Initiative Physique des Infinis (IPI), a research training program of the Idex SUPER at Sorbonne Universit\'e.
\correction{OA acknowledges funding support from CNES. Wavelet software was provided by C. Torrence and G. Compo, and is available at URL: http://atoc.colorado.edu/research/wavelets/. }

\appendix \label{sec:appendix}

\section{Model structures}
\subsection{Alfvén vortices}  \label{sec:Alfvén_vortices}
Alfvén vortices, introduced by \cite{1992Petviashvili}, are cylindrical magnetic structures with Alfv\'enic properties, i.e., with correlated (or anti-correlated) transverse magnetic and velocity perturbations, and with current aligned (or anti-aligned) with vorticity.
The model of Alfv\'en vortices is based on the reduced MHD equations \citep{Kadomtsev1974,Strauss1976}, 
where the principal assumptions are the perpendicular anisotropy in the wave-vector space, $k_{\perp}\gg k_{\parallel}$, and slow time variations, \oa{$d/dt \ll f_{ci}$}. 
Two main types of vortices are distinguished: monopolar and dipolar.

Let the axis $z$ be along the background magnetic field $\mathbf{B}_{0}$. The transverse magnetic field $\delta\mathbf{B}_{\perp}=\nabla A_z \times \mathbf{z}$ and velocity $\delta\mathbf{V}_{\perp}=\mathbf{z}\times \nabla\psi$ perturbations are expressed with the axial component of the vector potential $A_z$ and the velocity flux function $\psi$. The model assumes linear proportionality $\psi=\xi A_z$, or equivalently $\delta\mathbf{V}_{\perp}/V_A=\xi \,\delta\mathbf{B}_{\perp}/B_0$ (generalised Alfvén relation).

\subsubsection{Monopole Alfvén vortex}
\label{sec:Alfvén_monopoles}

A monopole Alfvén vortex is localized within the cylinder of the radius $a$, and the axis of the cylinder is aligned with $\mathbf{B}_{0}$. 
The model assumes that the total current inside $r<a$ is zero. If $\delta\mathbf{B}_{\perp}$ is continuous at $r=a$, it implies the condition $J_{1}(ka)=0$, where $J_{1}$ is the first order Bessel function. This defines the parameter $k$ for a given radius $a$. The monopole vortex solution writes (in \correction{dimensionless} units, see \cite{1992Petviashvili}):
  \begin{equation}   \label{eq:monopole}
    \begin{cases}
  A_{z}= A_{0}\,(J_{0}(kr)-J_{0}(ka)),  & r<a \\ 
  A_{z}=0, & r>a
    \end{cases}       
  \end{equation}
where $A_0$ is the monopole vortex amplitude and $J_0$ is the zero order Bessel function. 

A monopole Alfv\'en vortex in the plane perpendicular to its axis 
is shown in the top panel of Figure~\ref{fig:models}(a). 
The amplitude of the structure, $\delta B_{\perp}/B_{0}=0.5$, is taken to be comparable to the observations (see Figure \ref{fig:structure2}~(b)).

\subsubsection{Dipole Alfv\'en vortex}
\label{sec:Alfvén_dipoles}
As in \correction{the} case of the monopole vortex, the dipole vortex is a coherent structure \correction{localized} inside the cylinder of the radius $a$ and the \correction{generalized} Alfvén relation $\delta\mathbf{V}_{\perp}/V_A=\xi \,\delta\mathbf{B}_{\perp}/B_0$ is assumed.

The particular property of the dipole Alfv\'en vortex model is that \correction{its} axis can be inclined by a small angle $\theta$ with respect to the background magnetic field $B_z=B_0$. We define $\alpha=\tan(\theta)$.
Without restriction of generality, let the axis of the vortex be in the $(y,z)$-plane. 
If $\theta \ne 0$, the dipole vortex \correction{propagates} along $y$ with the speed $u\propto\alpha$. The continuity of $\delta\mathbf{B}_{\perp}$ at $r=a$ requires that the amplitude of the dipole vortex is not arbitrary, but defined by $\alpha$ and $k$.

In the reference frame moving with the vortex, the \correction{dimensionless} vector potential of the dipole vortex is \citep{1992Petviashvili,Alexandrova2008NPG}:
  \begin{equation}   \label{eq:dipole}
    \begin{cases}
  A_{z}= \frac{-2 \alpha}{k J_{0}(ka)} \, J_{1}(kr) \frac{x}{r}+\alpha\, x, &r<a\\
  A_{z}= \alpha x  \frac{a^{2}}{r^{2}}, &r>a
    \end{cases}       
  \end{equation}
A dipole Alfv\'en vortex is shown in Figure~\ref{fig:models}(b).

\subsection{Current sheets} \label{sec:Current_sheets}
Current sheets are planar coherent structures that separate the plasma with different magnetic field directions. 
Current sheets with large rotation angles across the sheet represent the boundaries of magnetic tubes, according to \citet{Bruno2001, Borovsky2008}. The population of current sheets with smaller rotation angles is much more numerous \citep{Borovsky2008}. They might be formed spontaneously as a result of the turbulent cascade, \citep[e.g.,][]{Veltri1999,Mangeney2001,Veltri2005,Servidio2008,2009Salem, Zhdankin2012, Greco2008, Greco2009, 2012Greco}.

MHD classification of current sheets \correction{includes} rotational (RDs) and tangential (TDs) discontinuities \citep[e.g.,][]{baumjohann97,2011Tsurutani}. 
A typical method to distinguish RD from TD is based on the \correction{normalized} change in magnetic field magnitude $\Delta B/B$ across the discontinuity (which is zero for RD) and the normal magnetic field component $B_{n}/B$ (which is zero for TD). \correction{To test the criterion $B_n/B=0$, the normal to the sheet must be accurately determined. Current sheets are planar structures, so $B_n=const$ (due to the divergence-free magnetic field). Therefore, the direction of the normal ($\mathbf{n}$) can be estimated using MVA, namely $\mathbf{n}=\mathbf{e_3}$. However, the error of this method is significant \citep{2001Horbury,2004Knetter,Wang2024}. Thus, the verification of the criterion by MVA in application to single-spacecraft data is not accurate enough to classify the current sheets in the solar wind as tangential or rotational.}
Observations showed that current sheets can combine \correction{the} properties of RDs and TDs, \correction{so classifying them as RDs or TDs might be an oversimplification} \citep[e.g.,][]{2006Neugebauer,2019Artemyev}.

\subsubsection{Rotational discontinuity} \label{sec:RD}
RDs are \correction{characterized} by the correlated rotation of magnetic field and velocity (Walen relation in case of the pressure isotropy: $ \mathbf{\delta B}/B_0 = \pm\mathbf{ \delta V}/V_A$), constant magnetic and plasma pressures across the sheet ($\Delta B/B = \Delta P/P=0$). 
Plasma on both sides of a RD is magnetically connected, i.e., $B_{n} \ne 0$. 

Let the normal to the current sheet $\mathbf{n}$ be along $\mathbf{e}_{x}$, $B_{n}$ and $B_{t}$ denote normal and tangential magnetic field components.
The condition $\mathbf{\nabla} \cdot \mathbf{B}=0$ implies $B_{x}=B_{n}=\text{constant}$. 
We use the same rotational discontinuity model as in \citet{Goodrich1991}, where the magnetic field rotates smoothly by an angle $\zeta(x/\ell)=\Delta\zeta /2 \tanh(x/\ell)$ with a total angle $\Delta \zeta$ across the RD with thickness $\ell$:
  \begin{equation}    \label{eq:RD}
  \begin{cases}
 	B_{x}(x) = B_{n} \\ 
 	B_{y}(x) = B_{t} \cos(\zeta(x/\ell))\\
	B_{z}(x) = B_{t} \sin(\zeta(x/\ell))
  \end{cases}
  \end{equation} 
We select $\Delta \zeta=120^{\circ}$ as in the example that we discuss in Section \ref{sec:Example3}. 
According to the statistical study of the current sheets from the first PSP perihelion, thin current sheets have smaller magnetic field rotation angles $\Delta \zeta \sim 21^{\circ} \times (\ell/d_i)^{0.32}$, where $\ell$ is the CS thickness and $d_i$ is the proton inertial length \citep{2022Lotekar}.
For rotational discontinuities with smaller $\Delta \zeta$ the polarization becomes closer to linear, i.e., closer to \correction{the} case of \correction{the} tangential discontinuity model discussed in Section \ref{sec:TD}. 
In terms of eigenvalues, $\lambda_{3}/\lambda_{2}$ increases 
while $\Delta \zeta$ decreases.
The selection of $\Delta \zeta=120^{\circ}$ corresponds to a high-amplitude RD. The 
RDs \correction{with small $\Delta \zeta$} cannot be distinguished from TD with the \correction{polarization} and MVA eigenvalue ratios.

\subsubsection{Tangential discontinuity} 
\label{sec:TD}
TDs separate two magnetically disconnected plasma regions, so \correction{the} normal component \correction{of the} magnetic field is zero, $B_x=B_{n}=0$. 
We use the Harris-like current sheet model, with a constant guide field $B_{z}=B_{0}$ (\cite{1962Harris}): 
  \begin{equation}   \label{eq:TD}
   \begin{cases}
   B_{x} = 0\\ 
   B_{y}= B_{t} \tanh(x/\ell)\\
   B_{z}= B_{0}
   \end{cases}       
  \end{equation}
In \correction{the} presence of the strong guide field $B_{0}\gg B_{t}$, the current density is quasi-parallel to \correction{the} magnetic field. So, the current sheet is quasi-force-free in accordance with what was found in observations \citep{2019Artemyev}.

\subsection{Magnetic holes} \label{sec:Magnetic_holes}
\correction{A} magnetic hole represents a localized magnetic field modulus decrease. 
MHD-scale magnetic holes (with cross-section widths ranging from $\sim 10\, \rho_{p}$ to $\sim 10^3 \rho_{p}$, where $\rho_{p}$ is the proton Larmor radius), are quite rare events:  at 1~AU the occurrence rate of 0.6 per day was observed by \cite{2007Stevens}. Closer to the Sun the occurrence rate\correction{ is higher}: 2.4 per day at 0.7 AU\correction{,} 3.4 per day at 0.3~AU \citep{2020Volwerk}.

MMS solar wind observations \citep{Wang2020} and kinetic simulations \citep{Roytershteyn2015, Haynes2015} have found magnetic holes at sub-ion scales. PIC simulations show that magnetic holes (defined as regions of magnetic field depression) tend to have cylindrical field-aligned geometry \citep{Roytershteyn2015, Haynes2015}.

We consider the magnetic hole model where the magnetic field direction does not change across the structure (linear magnetic hole). We suppose that the hole has cylindrical geometry and the axis is along $\mathbf{e}_{z}$. 
The radius of the hole is designated as $a$. 
  \begin{equation}     \label{eq:MH}
   \begin{cases}
   B_{x} = 0\\ 
   B_{y}= 0\\
   B_{z}= B_{0}  - \delta B_{\parallel}  
        / \cosh \left( (x/a)^2+(y/a)^2 \right) 
   \end{cases}       
  \end{equation}

\section{Examples of structures} \label{appendix:structures}


\subsection{Example 1} \label{sec:Example1}
\begin{figure*}[ht]
	\centering
	\includegraphics[width=1.02\linewidth]
 {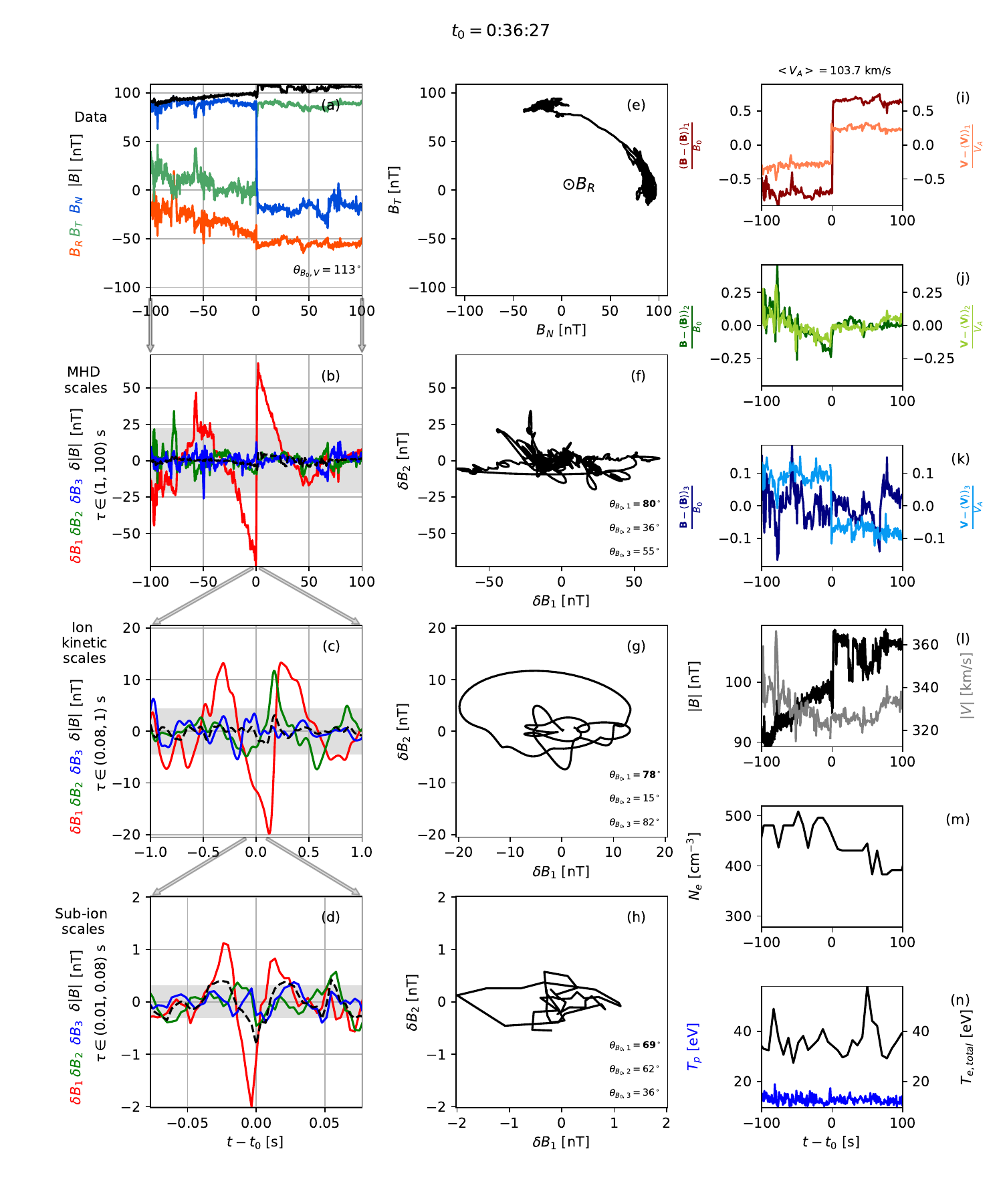}
 
\caption{ Example 1: A tangential discontinuity on MHD scales. The four panels on the left represent magnetic fluctuations near the central time $t_{0}$ = 0:36:27~UT.  (a) Original magnetic field in \correction{the} RTN reference frame, (b) bandpass filtered magnetic fluctuations at MHD, (c) ion and (d) sub-ion frequency ranges, see equations \ref{eq:frequency_ranges} and \ref{eq:timescale_ranges} in local MVA reference frame. 
(e-h) Corresponding \correction{polarization} plots. 
The orientation of MVA basis vectors 
$\{\mathbf{e}_i\}_{i=1,2,3}$, with respect to the background magnetic field $\mathbf{B}_0$ is provided in the legend with $\theta_{B_0,i}$ angles. 
Angles for well-defined MVA basis vectors are shown in bold.
Panels of the right column show different plasma parameters at MHD scales: (i-k) the correlation between $(\mathbf{B}-\langle \mathbf{B} \rangle)/B_{0}$ and $(\mathbf{V}-\langle \mathbf{V} \rangle)/V_{A}$, (l)
the magnetic field and velocity modulus ($|B|$ and $|V|$), (m) the electron density $N_{e}$, (n) proton and electron temperatures $T_{p}$, $T_{e,total}$.
\label{fig:structure1}}
\end{figure*}

The first event was observed on November~6, 2018, at $t_0=$00:36:27~UT. 
Figure~\ref{fig:structure1}(a) shows magnetic field data in the RTN reference frame during 200~s around $t_0$. 
Here, $B_T$ and $B_N$ components change \correction{the} sign in the center, 
\correction{the} magnetic field rotates by the angle $\Delta \zeta=80^{\circ}$. Thus, this is an example of a current sheet. PSP crosses this current sheet in $\Delta t\simeq 1.5$~s, so its thickness is about $V \Delta t \sim 450$~km.
The flow-to-field angle is $\Theta_{BV}=113^{\circ}$, so the PSP \correction{crosses} this structure under a quasi-perpendicular angle.
The polarization of the fluctuations in the plane $(B_N,B_T)$ is shown in panel (e). The out-of-plane $B_R$ is negative and nearly constant during the considered time interval, so this discontinuity is not at the edge of a switchback.

Figure~\ref{fig:structure1}(b) shows band-pass filtered MHD inertial range magnetic fluctuations $\delta\mathbf{B}_{MHD}$ during the same 200~s around $t_0$ in the MVA reference frame. 
The grey horizontal bands \correction{indicate} $\pm 2\sigma_{MHD}$ (two standard deviations of the random phase signal at MHD scales). 
The discontinuity in the center is due to the presence of the current sheet detected already in the raw data with the amplitude 
 $\delta B/B\simeq 1.4$. The shape of $\delta B_1$ (red line) around $t_0$ is due to the band-pass filtering or a current sheet shown above in panel~(a). The corresponding polarization (panel (f)) is nearly linear. In the legend, we indicate the angles between the corresponding mean field $\mathbf{B}_0$ and the MVA basis. The MVA basis vectors $\mathbf{e}_1$, $\mathbf{e}_2$ and $\mathbf{e}_3$ are well-defined if both eigenvalue ratios are small, $\lambda_2/\lambda_{1} < 0.3$ and  $\lambda_3/\lambda_{2} < 0.3$ \citep{1998Paschmann}. If only the first (second) of the ratios is below $0.3$, then only $\mathbf{e}_1$ ($\mathbf{e}_3$) is unambiguously defined. The angles with eigenvalue ratios below $0.3$ are shown in bold in the legend of the \correction{polarization} plane.
So, one can see a linear polarization, with the maximum variance direction $\mathbf{e_1}$  quasi-perpendicular to $\mathbf{B}_0$, $\Theta_{B,1}=80^{\circ}$. The intermediate $\mathbf{e}_2$ and minimum $\mathbf{e}_3$ variance directions are ill-defined.  

Figure~\ref{fig:structure1}(c) shows a zoom to the time interval of $\pm 1$~s around the same central time $t_0$.  The grey horizontal band indicates $\pm 2\sigma_{ion}$. For ion scales, the amplitude is significant, $\delta\mathbf{B}_{ion}/B_0 \sim 0.2$. The shape of $\delta\mathbf{B}_{ion}$ is not the filtering remnant of the current sheet, as shown in the Appendix~\ref{sec:Necessity_of_filtering}. 
The black dashed line shows the fluctuations of magnetic field modulus $\delta|\mathbf{B}|$, which are negligible.
Here, the local MVA frame is well defined. The minimum and maximum variation directions are perpendicular to the magnetic field. The elliptic polarization and the shape of magnetic fluctuations at ion scales resemble the crossing of a dipole vortex (shown in Figure~\ref{fig:models}(b)).
Thus, we observe a vortex-like structure at ion scales embedded in the current sheet at MHD scales.

Figure~\ref{fig:structure1}(d) shows the zoom-in to the time interval of $\pm0.07$~s around $t_0$.
\correction{High-amplitude fluctuations 
$\delta\mathbf{B}_{sub-ion}$ with respect to the noise level $\pm 2\sigma_{sub-ion}$ (grey band)} are well localized in time. \correction{The modulus of the magnetic field is fluctuating with a significant amplitude $\delta|B|/\delta B_{1}=0.4$ (black dashed line), so the fluctuations are compressible.}
\correction{This }is in agreement with a statistical increase  of compressibility at \correction{the} sub-ion range (see the spectrum of compressible fluctuations, Figure~\ref{fig:spectral}(b)).
\correction{The polarization is elliptic}, see panel~(h). The maximum MVA eigenvector is quasi-perpendicular to the background magnetic field $\theta_{B_0,1}=69^{\circ}$. These properties can be explained as \correction{the crossing of the} compressible vortex 
through its center \citep{Jovanovic2020}.

\begin{figure*}[ht]
	\centering
	\includegraphics[width=\linewidth]
{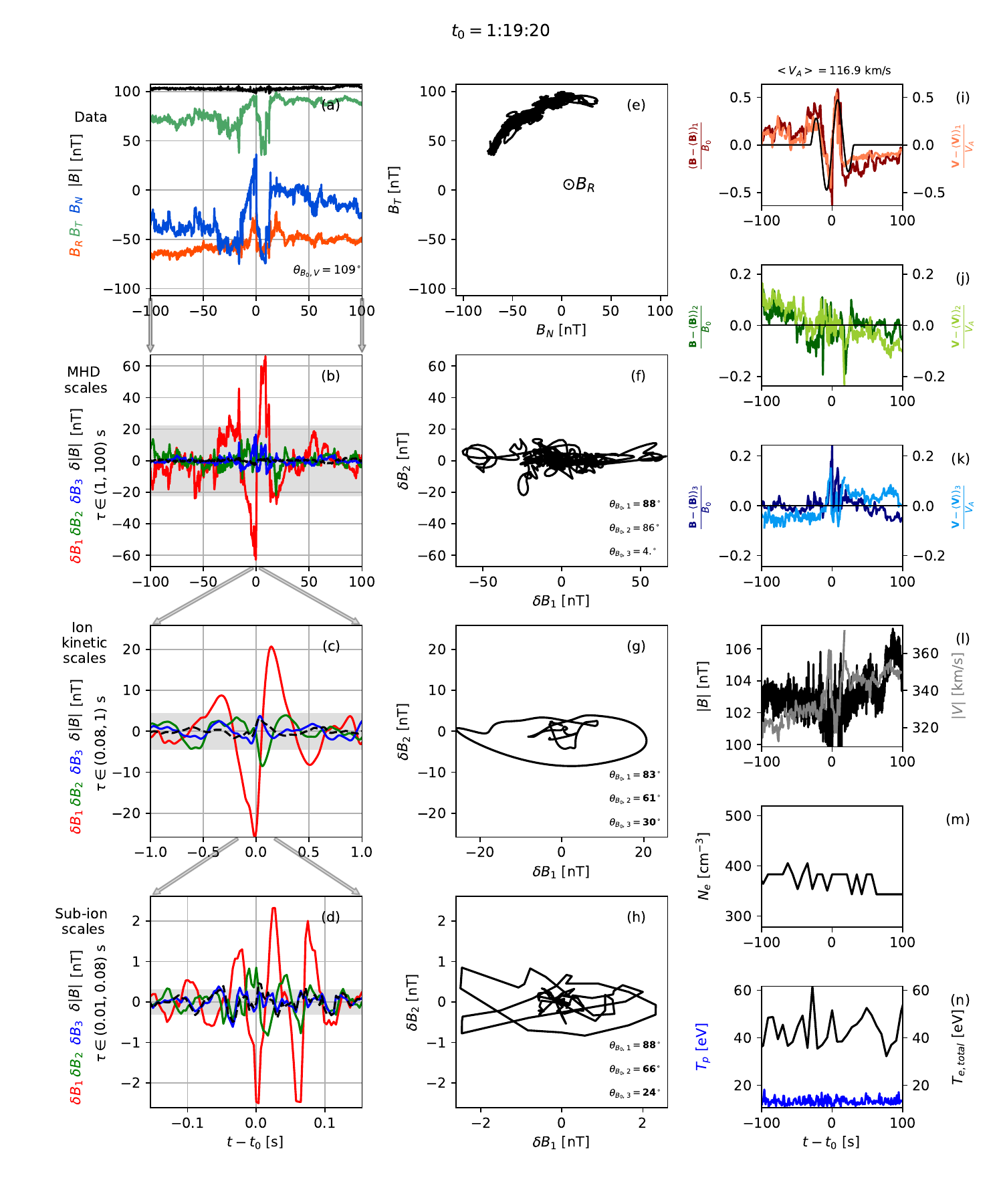}
\caption{Example 2: An Alfvén vortex embedded in a week current sheet on MHD scales. 
The format is the same as Figure \ref{fig:structure1}.
\label{fig:structure2}}
\end{figure*}

\begin{figure*}[ht]
	\centering
	\includegraphics[width=\linewidth]
 {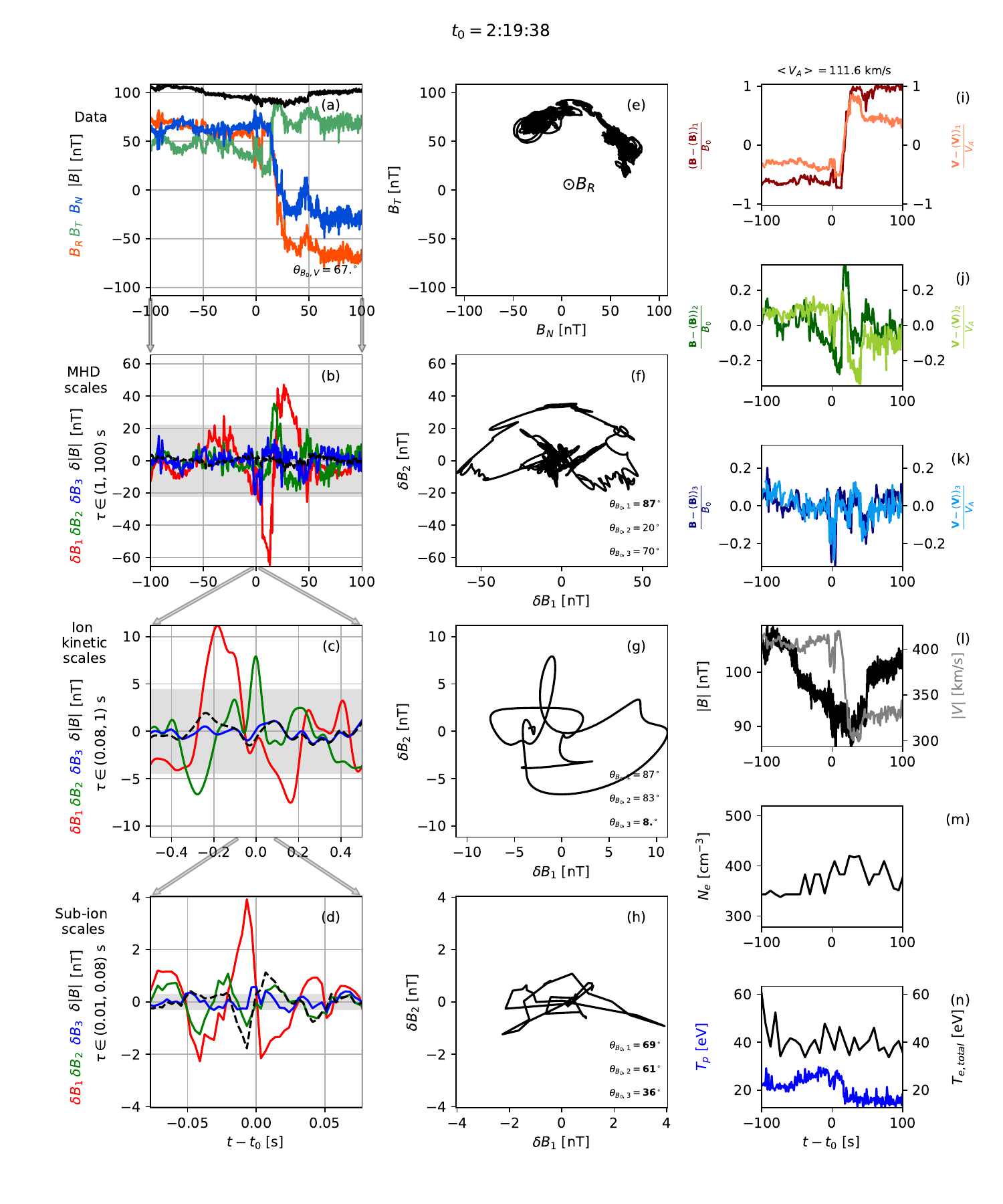}
\caption{Example 3: A rotational discontinuity (switchback boundary) nested in a larger magnetic depression region on MHD scales. The format is the same as Figure \ref{fig:structure1}.
\label{fig:structure3}}
\end{figure*}

Figure~\ref{fig:structure1}(i-k) show the magnetic field fluctuations \correction{normalized} by the background magnetic field $(\mathbf{B}-\langle \mathbf{B} \rangle)/B_{0}$, where $B_{0}=\langle |B| \rangle_{t-t_0 \in (-100,100) s}=100$ nT, and the proton velocity fluctuations $(\mathbf{V}-\langle \mathbf{V} \rangle)/V_{A}$ \correction{normalized} by the average Alfvén velocity $V_{A}=B_{0}/\sqrt{\mu_{0}N_{e}m_{p}}=104$ km/s. Both $(\mathbf{B}-\langle \mathbf{B} \rangle)/B_{0}$ and $(\mathbf{V}-\langle \mathbf{V} \rangle)/V_{A}$ are shown in magnetic field MVA reference frame calculated using $(\mathbf{B}-\langle \mathbf{B} \rangle)$ vector over the 200 seconds shown.  
Magnetic field and velocity variations across the sheet ($\Delta$) \correction{make the angle of}
$\alpha(\Delta \mathbf{B},\Delta \mathbf{V})=21^{\circ}$. 
Variations in $(\mathbf{B}-\langle \mathbf{B} \rangle)/B_{0}$ and $(\mathbf{V}-\langle \mathbf{V} \rangle)/V_{A}$ correlate, but the amplitudes are different $(|\Delta \mathbf{V}|/V_{A})_{1}=0.4 \cdot (|\Delta \mathbf{B}|/B_{0})_{1}$. 
Thus, the discontinuity does not fulfill the Walen relation $\Delta \mathbf{V}=\pm \Delta \mathbf{B}/\sqrt{4 \pi \rho}$ for rotational discontinuities.
In \correction{the} presence of pressure anisotropy, the density can change across the discontinuity, and the Walen relation is modified as follows \citep{Hudson1970, 2006Neugebauer}:
   \begin{eqnarray} 
\Delta(A\rho) &=& 0 \nonumber \\
\Delta \mathbf{V} &=& (\rho/\mu_0)^{1/2}A^{1/2} \Delta (\mathbf{B}/\rho)      \label{eq:DeltaV}
   \end{eqnarray}
where $A=1-\mu_0(p_{\parallel}-p_{\perp})/B^2$ is the anisotropy parameter. In the considered time interval $\beta_p<0.5$, so $A=1-4\pi(p_{\parallel}-p_{\perp})/B^2>1-\beta\sim 0.5$ implies $A^{1/2}>0.7$. However, we would need $A^{1/2} \sim 0.4$ to explain the observated relationship between $\Delta \mathbf{V}$ and $\Delta (\mathbf{B}/\rho)$ with anisotropy. 

In Figure~\ref{fig:structure1}(l-n) we see how magnetic field modulus $|\mathbf{B}|$, velocity modulus $|\mathbf{V}|$, electron density $N_e$, ion $T_{i}$ and electron $T_{e}$ temperatures change across the structure. 
Velocity and temperatures stay nearly constant. At the same time, $|\mathbf{B}|$ and $N_e$ are anti-correlated: while magnetic field increases by $\Delta|\mathbf{B}|=10$~nT, density decreases by $\Delta N_{e}\sim 50$~cm$^{-3}$. This is usually observed for convected structures in pressure balance.
The observed properties are typical for a tangential discontinuity, where magnetic field and density are not constant across the discontinuity. 

Another property \correction{that distinguishes} between RD and TD is the magnitude of the normal magnetic field component $B_n$. The divergence-free condition implies that $B_n$ must be constant in \correction{the} case of planar geometry. So, MVA minimum variance direction should represent normal to the magnetic sheet, and $B_{3}=B_{n}=$ constant. Next, tangential discontinuities have $B_{n}=0$, but in observations $B_3 \simeq 70$~nT. These results are obtained in the known limits of MVA since the MVA estimation of the normal to the sheet can differ from normal estimated from multi-spacecraft  methods \citep{2001Horbury,2004Knetter}.

So, to summarize, starting from the largest observed scales and up to the end of the inertial range, we observe a current sheet that can be interpreted \correction{as} a tangential discontinuity (TD). 
\correction{Ion} and sub-ion scale substructures are embedded in this discontinuity. Ion scale structure resembles the dipole Alfvén vortex model 
(see Section \ref{sec:models} and Figure \ref{fig:structure1} column (b)). 
Sub-ion scale structure might represent a compressible vortex \citep{Jovanovic2020}. 
A sketch describing this event is given in Figure \ref{fig:sketch}(a).

\subsection{Example 2}  \label{sec:Example2}

The second example is shown in Figure~\ref{fig:structure2} in the same format as the first event in Figure~\ref{fig:structure1}. The central time of the event is 01:19:20~UT. 
In panel (a), the raw magnetic field is shown in \correction{the} RTN reference frame. 

Panel (e) shows the polarization $B_T(B_N)$; out-of-plane $B_{R}$ does not change sign (this structure is not a switchback). 
The magnetic field deflects twice within the timescale of $\sim 80$~s. 
\correction{The magnetic field before crossing the structure and after are orientated differently:}
it is rotated by $\Delta \zeta=15^{\circ}$ (see Equation~(\ref{eq:RD})). 
This can be due to a weak ($|\Delta \mathbf{B}|/B_0 = (\mathbf{B}_{t=-100 s}-\mathbf{B}_{t=100 s})/B_0 \simeq 0.3$) rotational current sheet, since the ratio of velocity and magnetic field jumps satisfy the Walen relation |$\Delta \mathbf{V}|/V_{A}=1.03\cdot |\Delta \mathbf{B}|/B_{0}$. 

Magnetic fluctuations at the MHD scales $\delta \mathbf{B}_{MHD}$ are shown in Figure~\ref{fig:structure2}~(b) in the MVA reference frame.
The amplitude of the structure $\delta B_{1} \simeq 60$~nT (i.e., from peak to peak $\Delta B/B_0\sim 1.2$) well exceeds the level of incoherent signal $2\sigma_{MHD}=22$~nT.
The direction of the maximum eigenvector $\mathbf{e}_1$ is well-distinguished from intermediate ($\mathbf{e}_2$) and minimum ($\mathbf{e}_3$) directions since $\lambda_{2}/\lambda_{1}=0.06$, and it is perpendicular to the background magnetic field $\theta_{B_0,1}=88^{\circ}$.

The velocity and magnetic field fluctuations are well-correlated $\delta B/B_{0}=\xi  \delta V/ V_A$, see panels (i-k). 
 The proportionality coefficient is $\xi=0.86$ for magnetic fluctuations at MHD scales and $\xi=0.81$ in the raw data with the mean value subtracted for the same time interval. 
 The black lines in panels (i-k) show the fluctuations of the monopole Alfvén vortex model, along the central crossing. The observed time profiles of fluctuations agree well with the Alfven vortex model. The magnetic field vector before and after the vortex crossing is oriented differently: $|\mathbf{B}(t_0-70~\text{s})- \mathbf{B}(t_0+70~\text{s})|/B_0=(0.4,0.03,0.03)$ (in the MVA reference frame). Thus, the vortex is embedded in a \correction{relatively} weak current sheet.

In the plasma rest frame, this vortex propagates with a \correction{negligible} speed of $\simeq 3$ km/s (see the main text of the paper), so the Taylor hypothesis can be used to estimate its spatial scale. 
PSP trajectory crosses the structure in a plane nearly perpendicular to $\mathbf{B}_{0}$ ($\theta_{B_0,V}=71^{\circ}$). 
The diameter of the vortex can be estimated as $d=\theta_{B_0,V} \cdot V\Delta t \sim 2.4 \cdot 10^{4}$~km. 
Given that the mean ion inertial length is $d_i\simeq 11$~km and the mean ion Larmor radius is $\rho_i=5$~km, the vortex diameter $d \simeq 2 \cdot 10^{3} d_i = 5 \cdot 10^{3} \rho_i$. 
The variation of the magnetic field modulus is negligible $\delta |\mathbf{B}|/B_0=0.03$ as well as the variations of $N_{e}$, $T_{e}$ and $T_{p}$, see panels (l-n). 
The change of $|\mathbf{V}|$ (grey line in panel (l)) is due to the superposition of the velocity fluctuation of the Alfvén vortex on the bulk solar wind speed.

Figure~\ref{fig:structure2}(c) shows the ion scale magnetic fluctuations $\delta \mathbf{B}_{ion}$  located in the center of the MHD scale Alfvén vortex. 
The maximum amplitude of the fluctuation: $\max(|\delta B_{1}|)=24$~nT, as well as two secondary peaks on the left and right sides  exceed well the incoherent threshold $\pm 2\sigma_{ion-scale}$ shown in grey. 
The polarization is elliptical (panel (g)), and the maximum variance is perpendicular to the local field direction. These observed properties are in agreement with an Alfv\'en vortex crossing with a finite impact distance from its center.
The vortex was observed during $\Delta t=1.5$~s. The diameter of the vortex is $d=V\Delta t\sin(\Theta_{BV})=440$~km $=40~d_i=90~\rho_i$, where $\Delta t=1.5$~s is the \correction{crossing duration}, $V=330$~km/s, $\Theta_{BV}=70^{\circ}$, $d_i=11$~km, $\rho_i=5$~km.

Figure~\ref{fig:structure2}(d) shows the sub-ion fluctuations $\delta \mathbf{B}_{sub-ion}$, which are 10 times more intense than the incoherent threshold. They are quasi-transverse, $\theta_{B_0,1}=88^{\circ}$ and $\theta_{B_0,2}=64^{\circ}$, and weekly compressible, $\max(\delta B_{\parallel}) \approx\max(\delta B_{3})<0.2\max(\delta B_{1})$. The polarization is elliptical. The maximum variance is perpendicular to the local field, as in the case of ion and MHD scale structures.
The \correction{crossing duration} is $\Delta t \simeq 0.15$~s, i.e., the cross section scale is about 47~km, or 4$d_i$, or 9$\rho_i$. The fluctuations can be explained by the compressible Alfv\'en vortex model \citep{Jovanovic2020}.

In summary, in the example of Figure~\ref{fig:structure2}, in raw data we observe a weak current sheet with the thickness of the high amplitude MHD scale structure embedded in it. This current sheet is Alfv\'enic in nature that is the property of a rotational discontinuity. 
We can interpret the MHD structure embedded in the weak rotational current sheet as a monopole Alfv\'en vortex crossed close to its center. Within this monopole Alfv\'en vortex, we observe smaller vortices at ion and sub-ion scales.
Figure~\ref{fig:sketch}~(b) shows \correction{the sketch of the Example 2.}

\subsection{Example 3}  \label{sec:Example3}
Figure~\ref{fig:structure3} shows an example observed around $t_0=$ 2:19:38~UT. 
In panel (a), the center of the current sheet is observed at $t-t_{0}=20$~s, when the magnetic field rotates by the angle $\Delta \zeta=110^{\circ}$. 
$B_R$ changes sign across the sheet, so the sheet forms the boundary of a switchback, similarly to observations in \citep{2020Krasnoselskikh}. 

Figure~\ref{fig:structure3}(b) \correction{shows} MHD scale fluctuations in \correction{the} MVA reference frame. 
The amplitude of fluctuations associated with the discontinuity \correction{exceeds} the level of the incoherent signal (see grey horizontal band). Panels (e-f) show the corresponding polarizations.

The fluctuations are nearly Alfvénic in the vicinity of the discontinuity: $ (\mathbf{B}-\langle\mathbf{B}\rangle)_1 /B_{0} =1.25 \cdot (\mathbf{V}-\langle\mathbf{V}\rangle)_{1}/V_{A}$, when $t-t_{0} \in (0,40)$~s, see Figure \ref{fig:structure3}(i-k). 
But further away from the discontinuity $\Delta B/B_{0}$ and $\Delta V/V_{A}$ have different amplitudes: $\Delta B_{1}/B_{0}\approx 2.2 \cdot \Delta V_{1}/V_{A}$, \oa{where} $\Delta B_1=|B_1 (t_0-100 \text{\,s})-B_1 (t_0+100 \text{\,s})|$ and $\Delta V_1=|V_1 (t_0-100 \text{\,s})-V_1 (t_0+100 \text{\,s})|$. 
The magnetic field modulus decreases from 105~nT at the boundaries to 90~nT in the center (panel (l)). The duration of this magnetic cavity
is $\Delta t \simeq 100$~s, which corresponds to the scale of $l=\Delta t \cdot V  = 3.5\cdot 10^{4}$~km $=3\cdot 10^3 d_i=5 \cdot 10^3 \rho_i$. 
The density $N_{e}$, Figure \ref{fig:structure3}~(m), weakly increases across the discontinuity. The proton temperature $T_{p}$, Figure \ref{fig:structure3}~(n), is higher on the left side of the discontinuity than on the right, it \oa{has a local maximum around the}  discontinuity center  in contrast with a nearly uniform $T_{e}$.
There is an anti-correlation between the 
\correction{magnetic} field \correction{modulus} and plasma density $N_e$, indicating a possible pressure balance.  
The polarization of magnetic fluctuations is arch-like, \correction{which} is typical for rotational discontinuities \citep{1996Tsurutani, 2010Sonnerup, 2012Haaland, 2013Paschmann}.

Magnetic fluctuations at ion scales \oa{are shown in } Figure \ref{fig:structure3}~(c). 
The maximum and intermediate magnetic fluctuations ($\delta B_1$ and $\delta B_2$) are transverse and have nearly the same amplitude, the \correction{polarization} is close to elliptic, Figure~\ref{fig:structure3}~(g). 
The described properties (i.e. \correction{localized} transverse fluctuations with nearly elliptic \correction{polarization}) are consistent with the off-center monopole Alfv\'en vortex crossing, see Figure~\ref{fig:models}~(a).
The minimum MVA eigenvector $\textbf{e}_3$ is well-defined, because $\lambda_3/\lambda_2=0.06$ is small. From the model crossings (see Section \ref{sec:Monopoles}) we know that if the spacecraft is crossing the monopole Alfven vortex, $\textbf{e}_3$ is a good approximation for the axis of the vortex. We can conclude that the  axis of the vortex is nearly parallel to the background magnetic field, since $\theta_{B_0,3}=8^{\circ}$. 
 
The sub-ion scale structure, Figure~\ref{fig:structure3}(d), is well \correction{localized} in time and has a significantly compressible component, $\delta |B| \sim0.5 \, \delta B_{1}$. This high compressibility is a typical \correction{property} of structures at these scales. 
Such localized compressible magnetic fluctuations at sub-ion scales can be interpreted as the compressible Alfv\'en vortex \citep{Jovanovic2020}.

\subsection{Example 4}\label{sec:Example4}
\begin{figure*}[ht]
	\centering
	\includegraphics[width=\linewidth]
 {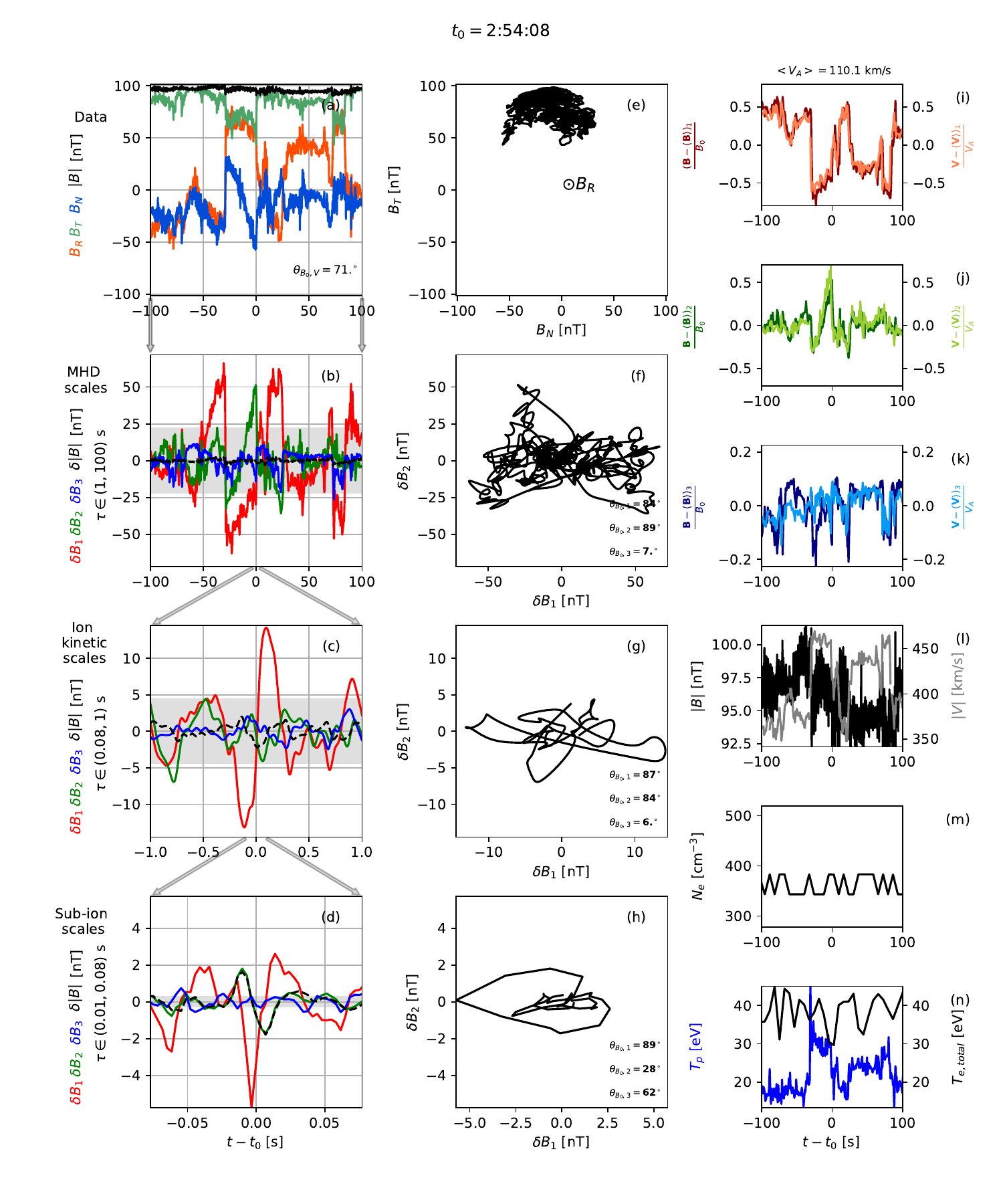}

\caption{Example 4: Two neighbour switchback structures.
The format is the same as Figure \ref{fig:structure1}.
\label{fig:structure4}}
\end{figure*}

Figure~\ref{fig:structure4} shows the details of the fourth example.
In panel~(a) the radial magnetic field $B_R$ is positive in two time intervals: $t-t_0\in(-30,0)$~s and $t-t_0\in(25,80)$~s. These are two \correction{neighboring} switchbacks.  
The crossing is nearly perpendicular to the mean background magnetic field $\mathbf{B_0}$ ($\theta_{BV}=71^{\circ}$), since $\mathbf{B_0}=\langle \mathbf{B} \rangle_{t-t_0\in(-100,100)~s}= (11, 84, -14)$~nT is directed \correction{mainly} along the tangential direction in the RTN reference frame and the solar wind bulk speed \oa{is} $\mathbf{V} =\langle \mathbf{V} \rangle_{t-t_0\in(-100,100)~s}= (400, 80, 0)$~km/s.

Figure~\ref{fig:structure4}(b) shows the MHD scale bandpass filtered magnetic fluctuations in \correction{the} MVA reference frame. 
We observe sharp discontinuities at the boundaries of both \correction{neighboring} switchbacks. 
The magnetic field and velocity fluctuations are well-correlated, the \correction{normalized} amplitudes of $\mathbf{B}-\langle\mathbf{B}\rangle)/B_0$ and $\mathbf{V}-\langle\mathbf{V}\rangle)/V_A$ are equal, see panels (i-k), so the fluctuations are Alfv\'enic. The density $N_e$ is constant across the structure, see panel (m). 
The proton temperature $T_p$ is increased by $\sim 30$\% in both \correction{neighboring} switchbacks compared to the value outside, see panel (n).

Similar to the previous examples, the embedded structures are observed \correction{near} the central time of the event $t-t_0=0$~s (at the right boundary of the first switchback). Fluctuations of ion scales are shown in the panel~(c). The polarization is close to linear, see panel~(g). The maximum variance direction is perpendicular to the local background magnetic field. The time profile of the corresponding component $\delta B_1$ is consistent with the monopole Alfv\'en vortex crossing through the center. 

Figure~\ref{fig:structure4}(d) shows the sub-ion fluctuations. The profile of the maximum variance component $\delta B_1$ has a peak. The intermediate component $\delta B_2$ coincide with $\delta|B|$, so the structure is compressible, with $\delta|B|/\max{|\delta B_1|}\sim 0.3$.  
\correction{We interpret this structure as a} compressible ion scale vortex \citep{Jovanovic2020}.

\subsection{Summary of detected structures}  \label{sec:Summary}
 We collected large statistics of coherent structures (Figure \ref{fig:intermittency_hist}). \correction{Some of these events} at MHD scales  represent isolated current sheets such as tangential and rotational current sheets, \correction{see} two examples shown in Section \ref{sec:Example1} and \ref{sec:Example3} respectively.  However, we found that current sheets are rare, most of the events can be interpreted as Alfv\'en vortices.
 The example in Section \ref{sec:Example2} (Figure \ref{fig:structure2}) is interpreted as the crossing of a monopole vortex along its center (embedded in a weak and large scale \correction{rotational} discontinuity). 
From the visual examination of $\sim 200$ events, we find that the embedded structures at ion and sub-ion scales are mostly Alfv\'en vortices, independently on the existence of a CS at large scales. From the examples shown here, in \correction{the} case of CS at large scales, the sub-ion vortices are compressible and in the case of the large scale Alfv\'en vortex, the small scale vortex is incompressible. 

\section{Necessity of band-pass filtering}\label{sec:Necessity_of_filtering}

\begin{figure}[ht!]
	\centering
	\includegraphics[width=.50\linewidth]{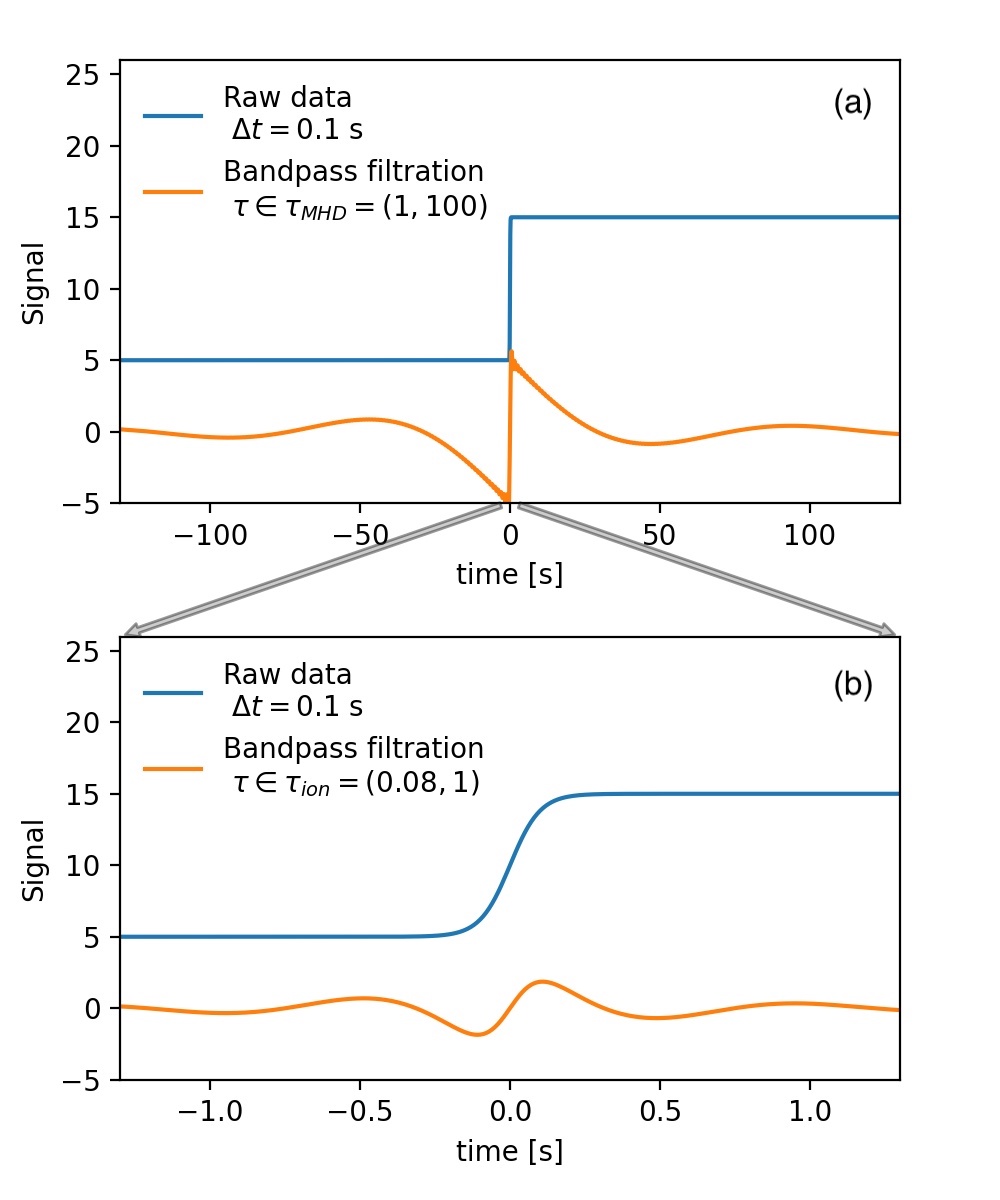}
\caption{Filtration of the current sheet. The temporal scale of the current sheet is $\Delta t=0.1$ s. The top panel (a) shows the raw data (blue line) and the bandpass filtered signal at MHD range of timescales $\tau \in \tau_{MHD}$ (orange). 
The bottom panel shows the zoom to the shorter time interval. The same raw data as in panel (a) is shown in blue. The orange line in panel (b) shows the bandpass filtration at $\tau \in \tau_{ion}$.} 
	\label{fig:appendix}
\end{figure}

At ion and sub-ion scales the amplitude of embedded substructures \correction{is} much smaller than the amplitude of MHD-scale fluctuations and the background magnetic field.
Therefore, filtration allows to remove the quasi-constant background, \correction{ which facilitates the} analysis of the fluctuations associated with substructures.
However, the filtration may introduce an ambiguity in the interpretation of the signal.
Let us consider the thin (i.e. $\ell \sim d_{i}$) current sheet, so that the crossing duration is $\Delta t=0.125$ s. 
Figure \ref{fig:appendix} shows the tangential magnetic field component of the current sheet (blue) and the result of the filtration (orange).

In Figure \ref{fig:appendix}(a) the filter frequency window   corresponds to the MHD inertial range, as defined in Section \ref{sec:spectrum} in Equation ~(\ref{eq:frequency_ranges}). Since the thickness of the sheet is small, $\Delta t/\text{min}(\tau_{MHD})=0.1$, the filtered signal has a steep jump of the same amplitude as the amplitude of the initial signal. However, unlike the initial signal, the filtered signal tends to 0 at the scale $|t|>\text{max}(\tau_{MHD})/2$ away from the discontinuity. Two  low amplitude local \correction{extremes} appear at $t=\pm 50$ s.

Figure \ref{fig:appendix}(b) shows the result of the filtration at ion scales. The magnetic field changes \correction{the} sign smoothly. If the thickness of an intense coherent structure is smaller than the minimum timescale of the MHD range, and if the 
classification method is based on the shape of the most intense filtered magnetic field component, then the CS filtration remnant at ion scales can be miss-classified as an embedded monopole Alfv\'en vortex crossed through its center.

In conclusion, filtering is necessary for the study of ion and subion scale structures because such structures have small amplitude compared to $B_0$, and they are poorly distinguishable in raw data.
However, filtering can significantly change the waveform, which complicates \correction{the} direct comparison of structures with models. For example, the current sheet does not look like a step function after filtering.
However, signal filtering has little effect on the \correction{polarization} of fluctuations. For example, if the spacecraft is crossing a tangential discontinuity (see Eq.~\ref{eq:TD}) with only one component of the magnetic field changed, i.e. with linear \correction{polarization}, then after filtering, the shape of the signal will change, but the \correction{polarization} will remain linear. \correction{Suppose} the spacecraft is crossing an Alfvén vortex and the measured \correction{polarization} is close to elliptical in raw data. \correction{In that case,} filtering will remove the quasi-constant background magnetic field, but the \correction{polarization} will remain elliptical. The \correction{polarization} is convenient to show the amplitude anisotropy of a 2D vector. But in general case the magnetic fluctuations are a 3D vector. So, the amplitude anisotropy can be \correction{characterized} by MVA eigenvalue relations.

\bibliographystyle{aasjournal}
\bibliography{References}

\end{document}